\documentclass[12pt]{article}
\usepackage{amssymb}
\usepackage{amsfonts}
\usepackage{mathrsfs}
\usepackage{latexsym}
\usepackage{amsmath}
\usepackage{graphics}
\usepackage{graphicx}
\usepackage{sectsty}

\sectionfont{\normalsize\bf}
\subsectionfont{\rm\normalsize}
\def\be{\begin{equation}}
\def\ee{\end{equation}}
\def\g{{\mathfrak g}}
\def\h{{\mathfrak h}}
\def\fP{{\mathfrak P}}
\def\fS{{\mathfrak S}}
\def\fR{{\mathfrak R}}
\def\fW{{\mathfrak W}}
\def\fA{{\mathfrak A}}
\def\fB{{\mathfrak B}}
\def\Zop{{\mathbb Z}}

\def\Cop{{\mathbb C}}

\def\Sop{{\mathbb S}}
\def\vac{|0\rangle}

\def\A{{\mathcal A}}

\def\Z{{\cal Z}}

\def\P{{\cal P}}
\def\T{{\cal T}}
\def\K{{\cal K}}
\def\U{{\cal U}}
\def\F{{\cal F}}
\def\Sc{{\cal S}}

\def\half{{\scriptstyle {1\over 2}}}
\def\twothirds{{\scriptstyle {2\over  3}}}

\def\sixth{{\scriptstyle {1\over  6}}}
\def\twelfth{{\scriptstyle {1\over  12}}}

\def\bmu{{\boldsymbol \mu}}

\def\ker{\hbox{Ker }}
\def\floop{{\hbox{\scriptsize loop}}}
\def\tree{{\hbox{\scriptsize tree}}}
\def\tr{\hbox{tr}}

\def\Res#1{\hbox{\vbox{\vskip8pt\hbox{Res}\vskip-6pt\hbox{$\scriptstyle#1=0$}\vskip-8pt}}\,\,}
\def\Resa#1{\hbox{\vbox{\vskip8pt\hbox{Res}\vskip-9pt\hbox{$\scriptstyle#1=0$}\vskip-8pt}}\,\,}
\numberwithin{equation}{section}

\begin{document}

\thispagestyle{empty}
\begin{flushright}
\end{flushright}
\baselineskip=16pt
\vspace{.5in}
{\Large
\begin{center}
{\bf Current Algebra on the Torus}
\end{center}}
\vskip 1.1cm
\begin{center}
{Louise Dolan}
\vskip5pt

\centerline{\em Department of Physics}
\centerline{\em University of North Carolina, Chapel Hill, NC 27599} 
\bigskip
\bigskip	
{Peter Goddard}
\vskip5pt

\centerline{\em Institute for Advanced Study}
\centerline{\em Princeton, NJ 08540, USA}
\end{center}
\bigskip
\bigskip

\abstract{\noindent 
We derive the N-point one-loop correlation functions for the
currents of an arbitrary affine Kac-Moody algebra. 
The one-loop amplitudes, which are elliptic
functions defined on the torus Riemann  surface,  are specified by
group invariant tensors and certain constant tau-dependent functions.
We compute the elliptic functions via a generating function, and explicitly 
construct the invariant tensor functions recursively in terms of Young tableaux.
The lowest tensors are related to the character formula of the representation of the
affine algebra.  
These general current algebra loop amplitudes provide a building block for 
open twistor string theory, among other applications.}
\bigskip

\setlength{\parindent}{0pt}
\setlength{\parskip}{6pt}

\vfill\eject
\section{Introduction}
Current algebra conformal field theory is often an important ingredient
to supply gauge symmetry in string theory. The tree level N-point
correlation functions of the
currents \cite{FZ} of an affine Kac-Moody Lie algebra \cite{Kac},  $\hat\g$, 
\be
[J^a_m,J^a_n]={f^{ab}}_cJ^c_{m+n}+\kappa^{ab}m\delta_{m,-n},
\label{alg}\ee
associated with a finite-dimensional algebra, $\g$,
are especially simple, and expressed as a sum over products of
differences, with the group tensors given by the level and structure constants
of the affine algebra.

The current correlators on the torus have more structure, but turn out to
be computable in terms of ellipitic functions, and specified by constant but
tau-dependent group invariant tensors.
Recursion relations for these correlation functions \cite{DG}
become tedious to evaluate for large numbers of currents.
In this paper we calculate the one-loop N-point current correlation functions
explicitly for an arbitrary Lie group, and describe their dependence on rather neat
combinations of Weierstrass functions and on group tensors given
in terms of the character of the representation.   

Loop calculations were considered for vertex operator algebras in \cite{Zhu,Zhutwo},
for particular constructions of current algebras in \cite{MT}, and
for particular Lie groups \cite{MM}.
Our general treatment of the affine current correlators is possible due to the 
simple holomorphic operator products of the currents. Loop correlation functions
for other fields related to current algebras tend to be less completely accessible, 
although widely studied 
\cite{EO}-\cite{TG}.

Our interest in the current algebra torus correlator was initially motivated
by its appearance in the gluon loop amplitude \cite{DG} for open twistor string theory
\cite{Berk},\cite{BW}.  The N-point torus current correlator should be helpful
to pursue perturbation theory there.  The twistor string \cite{Wit},
\cite{Berk},  and efforts to formulate it as a heterotic theory \cite{MS}, 
although mixing conformal supergravity with Yang-Mills, 
also provides an enticing framework for
a QCD string. Our analysis of current algebra on the torus provides a
fundamental building block that will have general applications.  

The plan of this paper is as follows. In section \ref{fermionictree}, 
we first use the representation of a current algebra as bilinear expressions 
in (Neveu-Schwarz) fermions to evaluate current algebra tree amplitudes. 
The expressions obtained involve the tensors formed from the traces of products of the real matrices representing $\g$ and can be described by a set of graphical rules that will be extended later in the paper to yield loop amplitudes. 
Although the tensors depend on the representation chosen, this dependence 
cancels out in the expressions for the tree amplitudes because these are 
determined by $\kappa^{ab}$. For a compact simple algebra, we can take 
$\kappa^{ab}=k\delta^{ab}$ and we can obtain the general tree amplitude by 
scaling terms in the result obtained for any 
given representation by appropriate powers of $k$.

Notwithstanding this,  in section \ref{sect:FZ}, we find it useful to give a more generally phrased version of the construction, due to Frenkel and Zhu (FZ) \cite{FZ}. This generalizes the traces of representation matrices to invariant $m$-th order tensors $\kappa_m$, satisfying conditions \eqref{kappaeq} and \eqref{kcycle}, which determine $\kappa_m$ in terms of $\kappa_{m-1}$ uniquely up to an arbitrary symmetric invariant tensor $\omega_m$. The successive freedoms, represented by the $\omega_n$, have no effect on the tree amplitudes constructed using the $\kappa_m$. We isolate a ``connected'' part of the tree amplitude, which possesses only simple poles and show that, like the full amplitude, this just depends on $\kappa_2=\kappa$, and so not on the $\omega_m$.  In section \ref{sect:Tensors} we give a proof that, given a suitable $\kappa_{m-1}$, there exists a $\kappa_m$ satisfying \eqref{kappaeq} and \eqref{kcycle}, and we give explicit formulae for the general $\kappa_3$ and $\kappa_4$. Our proof of the existence of $\kappa_n$ does not itself provide a convenient algorithmic construction and we give this in Appendix \ref{sect:constrkappa} using Young tableaux and the representation theory of the permutation group.

In section \ref{fermionloop}, we begin by computing the $n$-point one loop 
amplitude using the representation of the current algebra as   bilinears 
in Neveu-Schwarz fermions. The result is given by a modification of the 
graphical rules used in section \ref{fermionictree} to describe tree 
amplitudes. Similar rules describe two other versions of the loop: 
one in which we use Neveu-Schwarz fermions but also incorporate a factor of 
$(-1)^{N_b}$, were $N_b$ is the fermion number operator, into the trace 
defining the loop; and one where we use Ramond rather than Neveu-Schwarz 
fermions. These rules involve the tensors constructed from traces of 
representation matices, used in the fermionic construction of tree 
amplitudes. There nearly all the structure resulting from varying the 
representation, reflected in the `arbitrary' 
symmetric tensors $\omega_m$ 
coming into the FZ construction,  was irrelevant, but this is not so for 
the loops.

To approach the construction of the general one loop current algebra 
amplitude, we isolate a connected part of the amplitude in section 
\ref{connected}, which has only single poles, as we did for the 
tree amplitudes. The residues of this connected part for the $n$-point 
loop are specified in terms of the $(n-1)$-point loop and this means that 
the $n$ point loop is determined in this way up to a 
symmetric invariant $n$-th order tensor, $\omega_n(\tau)$, depending only on the torus modulus, $\tau$. In section \ref{structuretorus}, we first obtain general forms for the two- and three-point loops in terms of symmetric invariant tensors $\omega_2(\tau)$ and $\omega_3(\tau)$ and Weierstrass $\P$ and $\zeta$ functions.
The general form for the $n$-point loop is given by an adaptation of the rules for tree amplitudes, expressed in terms of Weierstrass $\sigma$ functions through
\be \nu^{-n}H_n=\prod_{j=1}^n{\sigma(\mu_j+\nu,\tau)\over\sigma(\nu,\tau)\sigma(\mu_j,\tau)}
=\sum_{m=0}^\infty H_{n,m}\nu^{m-n},
\label{ellipticH}\ee
which is elliptic as a function of $\nu$ and $\mu_1,\ldots\mu_n$, provided that $\sum_{j=1}^n\mu_j=0$, and in terms of $n$-th order invariant tensor functions of $\tau$, $\kappa_{n,m}(\tau)$, with $n\geq m\geq 2$, defined inductively by \eqref{kappatau1} and \eqref{kappatau2} (which are similar to  \eqref{kappaeq} and \eqref{kcycle}), starting from invariant symmetric tensors
$\kappa_{n,0}(\tau)=\omega_n(\tau)$. In appendix \ref{sect:PropHn} we discuss properties of the functions $H_{n,m}$ and in appendix \ref{sect:formulae} we show how the general results of this section relate to those previously obtained in \cite{DG} for two-, three- and four-point loops.

The symmetric tensors $\omega_n$, irrelevant in the construction of tree amplitudes in section \ref{CAtree}, provide the extra structure necessary for the construction of the one-loop amplitudes. They are not arbitrary but can be determined in terms of traces of zero modes of the currents, 
$\tr\left(J^{a_{i_1}}_0J^{a_{i_2}}_0\ldots J^{a_{i_n}}_0w^{L_0}\right)$,
 symmetrized over the indices $a_j$.  In \ref{sect:recurrence}, we establish recurrence relations relating the traces over symmetrized products of currents, in terms of which the $\omega_n(\tau)$ are initially defined,  to symmetrized traces of their zero modes, showing how this works out in detail for $n=2,3$ and $4$. More precisely, $\omega_n(\tau)$ are defined in terms of the connected parts of the symmetrized traces of currents and, in \ref{connectzero}, we use the recurrence relations to determine
 $\omega_n(\tau)$ in terms of the connected part of the symmetrized trace of zero modes. 
 
 Then, in section \ref{sect:zeromodesch}, we show how the symmetrized traces of zero modes of the currents can be determined in terms of  
 \be\chi(\theta,\tau)=\tr\left(e^{i H\cdot\theta} w^{L_0}\right),\label{chidef0}\ee
the character of the representation of $\hat\g$ provided by the space of 
states of the theory.  While 
the analysis up to this point has  not 
made any assumptions about the Lie algebra $\g$, in this section we assume 
that it is compact and, for ease of exposition, take it to be simple. 
The method depends on using the Harish-Chandra isomorphism  of the
center of the enveloping algebra of $\g$, that is the ring of Casimir operators of $\g$, onto the polynomials in $H$ invariant under the action of the Weyl group, $W_\g$ of $\g$. 
 
Section \ref{conclusions} provides a summary of our results.
 
\vfil\eject

\section{Current Algebra Trees}
\label{CAtree}
\subsection{\sl{Current Algebra and the Fermionic Tree Construction}}
\label{fermionictree}

We consider a conformal field theory containing the affine algebra, $\hat\g$, given by \eqref{alg}, where
$m, n$  are 
integers and ${f^{ab}}_c$ are the  structure constants of $\g$ and $\kappa^{ab}$ is a symmetric tensor invariant with respect to $\g$. [If the generators of the algebra satisfy the hermiticity condition ${J^a_n}^\dagger=J^a_{-n}$,
${f^{ab}}_c$ is pure imaginary and $\kappa^{ab}$ is real.] For a general introductory review see \cite{GO}.

We consider evaluating the 
vacuum expectation value 
\be\A_\tree^{a_1a_2\ldots a_n}(z_1,z_2,\ldots,z_n)
=\langle 0|J^{a_1}(z_1)J^{a_2}(z_2)\ldots J^{a_n}(z_n)|0\rangle,\label{ourtree}\ee
where 
\be J^a(z)=\sum_n J^a_nz^{-n-1}, \qquad J^a_n\vac=0, \quad n\geq  0,
\qquad (J^a_n)^\dagger=J^a_{-n}.\label{conff}\ee
The currents $J^a(z)$ satisfy the operator product expansion 
\be
J^a(z_1)J^b(z_2)\sim{\kappa^{ab}\over (z_1-z_2)^2}+ {{f^{ab}}_cJ^c(z_2)\over z_1-z_2}
\label{ope}
\ee
and the tree amplitudes satisfy the asymptotic condition 
\be\A_\tree^{a_1a_2\ldots a_n}(z_1,z_2,\ldots,z_n)
={\cal O}(z_j^{-2})\quad\hbox{as}\quad z_j\rightarrow\infty,\label{treelim}\ee
because in this limit 
$\langle 0|J^a(z)\sim \langle 0|J^a_1z^{-2}.$

Because of the locality of the currents $J^a(z)$ relative to one another, the tree amplitude \eqref{ourtree}
is symmetric under simultaneous permutations of the $z_i$ and $a_i$,
\be\A_\tree^{a_{\varrho(1)}a_{\varrho(2)}\ldots a_{\varrho(n)}}(z_{\varrho(1)},z_{\varrho(2)},\ldots,z_{\varrho(n)})
=\A_\tree^{a_1a_2\ldots a_n}(z_1,z_2,\ldots,z_n),\label{treeperm}\ee
where $\varrho\in\fS_n$, the group of permutations on $n$ objects

The condition \eqref{ope} gives all the singularities of the $n$-point function in terms of $(n-1)$- and $(n-2)$-point functions. Thus, given \eqref{treelim}, using Cauchy's Theorem, we can inductively calculate the $n$-point function for any $n$ starting from the two-point function,
\be\langle 0|J^a(z_1)J^b(z_2)\vac = {\kappa^{ab}\over (z_1-z_2)^2},\label{prop}\ee
{\it i.e.} the $n$-point function is determined by the invariant symmetric tensor $\kappa^{ab}$.
A general prescription for doing this has been given by Frenkel and Zhu \cite{FZ}, which we shall discuss in section \ref{sect:FZ}, but first we shall note the explicit calculation when $J^a(z)$ is given as a bilinear in fermionic oscillators. 

Given a representation $J^a_0\mapsto t^a= iM^a$ of $\g$, where the $M^a$ are $N$-dimensional real antisymmetric matrices satisfying
\be[M^a,M^b]=-i{f^{ab}}_cM^c,\label{Malg}\ee
we can represent $J^a(z)$ as a bilinear in Neveu-Schwarz fermionic fields,
\be J^a(z)=\sum_{n\in\Zop}J^a_nz^{-n-1}={i\over 2}M^a_{ij}b^i(z)b^j(z)\label{Mbb}\ee
where
\be b^i(z)=\sum_{r\in\Zop+\half}b^i_rz^{-r-\half},\qquad \{b_r^i,b_s^j\}=\delta_{r,-s}\delta^{ij},\qquad
b_r^i\vac=0, \quad r>0,\qquad 1\leq i,j\leq N.\label{balg}\ee
Then $J^a_n$ satisfies \eqref{alg} with
\be\kappa^{ab}=-\half\tr(M^aM^b)=\half\tr(t^at^b).\label{kMM}\ee
Note
\be
b^i(z_1)b^j(z_2)=:b^i(z_1)b^j(z_2):\,+\,{\delta^{ij}\over z_1-z_2},
\label{bprop}\ee
with the usual definition of normal ordering.

Using Wick's theorem, we can evaluate the tree amplitude \eqref{ourtree} and describe the result as follows. The $n$-point function can be written as a sum over permutations $\varrho\in\fS_n$ with no fixed point. Each such permutation
can be written as a product of cycles, $\varrho=\xi_1\xi_2\ldots\xi_r$ and we associate to $\varrho$ a product
$F_\varrho=(-1)^rf_{\xi_1}f_{\xi_2}\ldots f_{\xi_r}$, where the function $f_\xi$ is associated with the cycle $\xi=(i_1,i_2\ldots i_m)$, defined by 
\be
f_\xi={\half\tr(t^{a_1}t^{a_2}\ldots t^{a_m})\over
(z_{i_1}-z_{i_2})(z_{i_2}-z_{i_3})\ldots (z_{i_m}-z_{i_1})}.\label{bcycle}\ee
The $n$-point tree amplitude is then constructed as the sum of these products over the permutations $\varrho\in \fS'_n$, the subset of $\fS_n$ with no fixed points,
\be
\A_\tree^{a_1a_2\ldots a_n}(z_1,z_2,\ldots,z_n)
=\sum_{\varrho\in\fS'_n} F^{a_1a_2\ldots a_n}_\varrho(z_1,z_2,\ldots,z_n).
\label{treep}\ee

\subsection{\sl The Frenkel-Zhu  Construction}
\label{sect:FZ}

Frenkel and Zhu have shown how the fermionic construction of the last section can be modified to give the general construction for the tree amplitude \eqref{ourtree}.  Again the $n$-point function \eqref{ourtree} is written as a sum over permutations with no fixed point, $\varrho=\xi_1\xi_2\ldots\xi_r$, written as a product of cycles, with which is associated
$F_\varrho=(-1)^rf_{\xi_1}f_{\xi_2}\ldots f_{\xi_r}$, where now 
\be
f_\xi={\kappa_m^{a_{i_1}a_{i_2}\ldots a_{i_m}}\over
(z_{i_1}-z_{i_2})(z_{i_2}-z_{i_3})\ldots (z_{i_m}-z_{i_1})},\label{cycle}\ee
and the $m$-order tensors $\kappa_m$ are defined inductively by the conditions 
\be\kappa_m^{a_1a_2a_3\ldots a_m}-\kappa_m^{a_2a_1a_3\ldots a_m}={f^{a_1a_2}}_b\kappa_{m-1}^{ba_3\ldots a_m},
\label{kappaeq}\ee
and
\be\kappa_m^{a_1a_2a_3\ldots a_m}=\kappa_m^{a_2a_3\ldots a_ma_1}.\label{kcycle}\ee
The $n$-point tree amplitude is then constructed as in \eqref{treep}.

A graphical way of describing the Frenkel-Zhu construction (or the fermionic construction) is to say that the $n$-point  tree amplitude is given by summing over all graphs with $n$ vertices where the  vertices carry the labels $1,2,\ldots, n,$ and each vertex is connected by directed lines to other vertices, one of the lines at each vertex pointing towards it and one away from it. Then each graph consists of a number of directed ``loops'' or cycles, $\xi=(i_1,i_2\ldots i_m)$, with which we associate the expression \eqref{cycle} and the expression associated with the whole graph is the product of the expressions for the various cycles multiplied by a factor of $-1$ for each cycle. 

As is implied by comparing \eqref{bcycle} and \eqref{cycle}, a solution to the conditions \eqref{kappaeq} and \eqref{kcycle} can be constructed by setting $\kappa_m^{a_1a_2a_3\ldots a_m}=\tr(t^{a_1}t^{a_2}\ldots t^{a_m})$ or, more generally,
\be\kappa_m^{a_1a_2\ldots a_m}=\tr(Kt^{a_1}t^{a_2}\ldots t^{a_m}),\label{kappaKtr}\ee
where $t^a$ is any finite-dimensional representation of $\g$, {\it i.e.}
\be [t^a,t^b]={f^{ab}}_ct^c,\label{galg}\ee
and $K$ is any matrix commuting with all the $t^a$, {\it i.e.} invariant under the action of $\g$.
$K$ is to be chosen so that $\kappa_2^{ab}=\tr(Kt^at^b)=\kappa^{ab}$ as in \eqref{alg}, which can be done for any invariant tensor $\kappa^{ab}$ if $\g$ is compact and $t^a$ a faithful representation.

It is straightforward to verify that \eqref{treep}
has the singularity structure implied by the operator product expansion \eqref{ope}, provided that $\kappa_m$ satisfies \eqref{kappaeq} and \eqref{kcycle}, and satisfies the asymptotic condition \eqref{treelim}, and thus is inductively determined by Cauchy's Theorem, given the two-point function \eqref{prop}. Thus, it  does not depend on the choice of $\kappa_m$ satisfying \eqref{kappaeq} and \eqref{kcycle}, apart from through $\kappa_2=\kappa$. 
(In particular, although different choices of representation $t^a$ result in different tensors $\kappa_m$, as defined through \eqref{kappaKtr}, these differences cancel out in \eqref{cycle}, apart from dependence on $\kappa_2$.)
In fact, the stronger statement holds that the connected parts, that is the sums of \eqref{cycle} over permutations of $(i_1,i_2\ldots i_m)$, only depend on the $\kappa$'s through
$\kappa_2$. This is expressed in the following Proposition:

{\bf Proposition 1.} If $\g$ is a Lie algebra and the tensors $\kappa_m^{a_1a_2\ldots a_m}$, where $1\leq a_j\leq \dim \g$, are defined for $m\leq N$, and satisfy 
\be\kappa_m^{a_1a_2a_3\ldots a_m}-\kappa_m^{a_2a_1a_3\ldots a_m}={f^{a_1a_2}}_b\kappa_{m-1}^{ba_3\ldots a_m},
\label{kappaeq2}\ee
and
\be\kappa_m^{a_1a_2a_3\ldots a_m}=\kappa_m^{a_2a_3\ldots a_ma_1},\label{kcycle2}\ee
where ${f^{ab}}_c$ are the structure constants of $\g$, then the tensor functions 
\begin{align}
\A_{\tree, C}^{a_1a_2\ldots a_m}(z_1,z_2,\ldots, z_m)
&={1\over m}\sum_{\varrho\in \fS_m}f_{(\varrho(1),\varrho(2),\ldots,\varrho(m))}\label{Atreef1}\\
&={1\over m}\sum_{\varrho\in \fS_m}
{\kappa_m^{a_{\varrho(1)}a_{\varrho(2)}\ldots a_{\varrho(m)}}\over
(z_{\varrho(1)}-z_{\varrho(2)})(z_{\varrho(2)}-z_{\varrho(3)})
\ldots (z_{\varrho(m)}-z_{\varrho(1)})}\cr
&=\sum_{\varrho\in \fS_{m-1}}
{\kappa_m^{a_{\varrho(1)}a_{\varrho(2)}\ldots a_{\varrho(m-1)}a_m}\over
(z_{\varrho(1)}-z_{\varrho(2)})(z_{\varrho(2)}-z_{\varrho(3)})\ldots 
(z_{\varrho(m-1)}-z_m)(z_m-z_{\varrho(1)})}\cr
\label{Atreef2}\end{align}
depend on the $\kappa_m$ only through  $\kappa_2$.

{\sl Proof of the Proposition:} The result follows from Cauchy's Theorem because the functions $F$ defined by \eqref{Atreef2} satisfy
$$\A_{\tree, C}^{a_1a_2\ldots a_m}(z_1,z_2,\ldots, z_m)={\cal O}(z_1^{-2}),\quad\hbox{as}\quad z_1\rightarrow\infty$$
and
$$\A_{\tree, C}^{a_1a_2a_3\ldots a_m}(z_1,z_2,\ldots, z_m)\sim
{{f^{a_1a_2}}_b\over z_1-z_2} \A_{\tree, C}^{ba_3\ldots a_m}(z_2,\ldots, z_m)\quad\hbox{as}\quad
z_1\rightarrow z_2,$$
and so can be calculated inductively from 
$$\A_{\tree, C}^{ab}(z_1,z_2)=
{\kappa^{ab}\over (z_1-z_2)^2}.
$$

\subsection{\sl{The Tensors $\kappa_n$.}}
\label{sect:Tensors}

The conditions 
\be\kappa_n^{a_1a_2a_3\ldots a_n}-\kappa_n^{a_2a_1a_3\ldots a_n}={f^{a_1a_2}}_b\kappa_{n-1}^{ba_3\ldots a_n},
\label{kappaeq3}\ee
and
\be\kappa_n^{a_1a_2a_3\ldots a_n}=\kappa_n^{a_2a_3\ldots a_na_1}.\label{kcycle3}\ee
are sufficient to ensure that the amplitudes defined by  \eqref{treep} 
with $F_\varrho=(-1)^rf_{\xi_1}f_{\xi_2}\ldots f_{\xi_r}$, 
where $f_\xi$ is given by  \eqref{cycle}, depend 
on the $\kappa_m$ only through $\kappa_2$. However, $\kappa_2$ does not
uniquely determine  $\kappa_m$ through \eqref{kappaeq3} and \eqref{kcycle3}.
In this section, we shall discuss the existence and uniqueness of solutions to these equations. Although, the arbitrariness in $\kappa_m$, given $\kappa_2$, is not relevant for tree amplitudes, we shall see in \S\ref{sect:Loop} that this freedom is very relevant for the construction of the one-loop amplitudes.

The conditions \eqref{kappaeq3} and \eqref{kcycle3} have some immediate consequences. First, if $\kappa_{n-1}$
satisfies  \eqref{kappaeq3} for some $\kappa_n$ which also satisfies \eqref{kcycle3},  then 
$\kappa_{n-1}$ is invariant because
$$\sum_{j=1}^n{f^{ba_j}}_c\kappa^{a_1\ldots a_{j-1}ca_{j+1}\ldots a_n}
=\sum_{j=1}^n(\kappa^{a_1\ldots a_{j-1}ba_ja_{j+1}\ldots a_n}
-\kappa^{a_1\ldots a_{j-1}a_jba_{j+1}\ldots a_n})=0$$
using \eqref{kappaeq3} and then \eqref{kcycle3}. Thus for \eqref{kappaeq3} and \eqref{kcycle3} to have a solution for a given $\kappa_{n-1}$ then this tensor must be invariant.

Second, if $\kappa_n^{a_1a_2a_3\ldots a_n}$ and $\tilde\kappa_n^{a_1a_2a_3\ldots a_n}$ both satisfy \eqref{kappaeq3} with the same $\kappa_{n-1}^{a_1a_2a_3\ldots a_{n-1}}$ and both satisfy the cyclic property \eqref{kcycle3}, then the difference
$$\omega_n^{a_1a_2a_3\ldots a_n}=\tilde\kappa_n^{a_1a_2a_3\ldots a_n}-\kappa_n^{a_1a_2a_3\ldots a_n}$$
is cyclically symmetric and satisfies
$$\omega_n^{a_1a_2a_3\ldots a_n}=\omega_n^{a_2a_1a_3\ldots a_n}.$$
These two symmetries generate the whole of $\fS_n$ so that $\omega_n$
must be a symmetric tensor. Conversely, if $\omega_n$ is symmetric, it follows
that $\tilde\kappa_n=\kappa_n+
\omega_n$ satisfies \eqref{kappaeq3} and \eqref{kcycle3} if $\kappa_n$ does.
So $\kappa_{n-1}$ defines $\kappa_n$ through  \eqref{kappaeq3} and \eqref{kcycle3}, assuming a solution exists, up to a symmetric tensor $\omega_n$. We establish the existence of the solution in the following Proposition:

{\bf Proposition 2.} If $\g$ is a Lie algebra, define inductively the spaces $\K_n$ to consist of the invariant $n$-th order tensors  $\kappa_n^{a_1a_2\ldots a_n}$, where $1\leq a_j\leq \dim \g$, such that 
\be\kappa_n^{a_1a_2a_3\ldots a_n}-\kappa_n^{a_2a_1a_3\ldots a_n}={f^{a_1a_2}}_b\kappa_{n-1}^{ba_3\ldots a_n},
\label{kappaeq4}\ee
for some $\kappa_{n-1}\in\K_{n-1},$
where ${f^{ab}}_c$ are the structure constants of $\g$, 
and
\be\kappa_n^{a_1a_2a_3\ldots a_n}=\kappa_n^{a_2a_3\ldots a_na_1},\label{kcycle4}\ee
with $\K_0=\{0\}$.  Then, for each $\kappa_{n-1}\in\K_{n-1}$,  there exists a  $\kappa_n$ satisfying \eqref{kappaeq4} and \eqref{kcycle4} that is unique up to the addition of a symmetric invariant tensor  $\omega_n$. The solution can be uniquely specified by requiring that it be orthogonal to all symmetric $n$-th order tensors.

{\sl Proof of the Proposition:} We define the action of $\varrho\in \fS_n$ on $n$-th order tensors $\tau_n$ by
$$
(\varrho \tau_n)^{a_1a_2\ldots a_n}=\tau_n^{a_{\varrho^{-1}(1)}a_{\varrho^{-1}(2)}\ldots a_{\varrho^{-1}(n)}}
$$
so that this provides a representation of $\fS_n$ on $n$-th order tensors: $(\varrho\sigma)\tau_n =\varrho(\sigma\tau_n)$. For any $n$-th order tensor $\tau_n$ write
\be\eta(\varrho, \tau_n) = \tau_n-\varrho \tau_n;\label{tau2}\ee
then we can write
\be\tau_n={1\over n!} \sum_{\varrho\in \fS_n}\eta(\varrho,\tau_n)+\omega_n,\label{tau1}\ee
where
\be\omega_n ={1\over n!} \sum_{\varrho\in \fS_n} \varrho \tau_n,\label{omega1}\ee
is the symmetrization of the tensor $\tau_n$. If $\tau_n$ is invariant, $\eta(\varrho,\tau_n)$ is also invariant. Then, if $\kappa_n\in\K_n$,
\be{1\over n!} \sum_{\varrho\in \fS_n}\eta(\varrho,\kappa_n)\label{eta1}\ee
is also in $\K_n$ and satisfies \eqref{kappaeq4} for the same $\kappa_{n-1}\in\K_{n-1}$; further, it is orthogonal to any symmetric tensor. It is clear from \eqref{tau1} that, taking $\tau_n=\kappa_n\in\K_n$,
$\kappa_n$ is orthogonal to all symmetric tensors only if $\omega_n$, defined as in \eqref{omega1}, vanishes. Thus, \eqref{eta1} is the unique solution to \eqref{kappaeq4} and  \eqref{kcycle4} for the given $\kappa_{n-1}$, with this property.

We now proceed to use the expression \eqref{eta1} to show there exists a solution to \eqref{kappaeq4} and  \eqref{kcycle4} for a given $\kappa_{n-1}\in\K_{n-1}$. From \eqref{tau2},
\be\eta(\varrho_1\varrho_2, \kappa_n)= \kappa_n-\varrho_1\varrho_2 \kappa_n=\eta(\varrho_1, \kappa_n)+\varrho_1\eta(\varrho_2.\kappa_n)\label{eta2}\ee
and, so,
\be\eta(\varrho_1\ldots\varrho_k, \kappa_n)=\sum_{j=1}^k\varrho_1\ldots\varrho_{j-1}\eta(\varrho_j,\kappa_n).\label{eta3}\ee
We can use this to give a formula for a given $\kappa_n$, in terms of $\kappa_{n-1}\in\K_{n-1}$, by expressing each $\varrho\in \fS_n$ as a product of transpositions of adjacent indices and then using \eqref{kappaeq4}  and  \eqref{kcycle4}. However, such expressions are not unique, so we need to address this by first working in the free group, $\tilde \fS_n$, generated by these transpositions, defining a function $\tilde\phi: \tilde\fS_n\rightarrow\T_n$, the space of $n$-th order invariant tensors, for each $\kappa_{n-1}\in\K_{n-1}$, and then checking that we can impose the appropriate relations to obtain a definition for $\varrho\in\fS_n$. In this way, we will obtain an $n$-th order tensor $\phi(\varrho,\kappa_{n-1})$,  $\varrho\in\fS_n, \kappa_{n-1}\in\K_{n-1}$, which will provide the desired element $\phi(\kappa_{n-1})\in\K_n$ on averaging over $\varrho\in\fS_n$.  To show this, we finally show that $\phi(\kappa_{n-1})$ satisfies \eqref{kappaeq4} and  \eqref{kcycle4}.

$\fS_n$ is generated by transpositions $\{\sigma_i : 1\leq i\leq n\}$ where
\be\sigma_i(i)=i+1,\quad \sigma_i(i+1)=i,\quad \sigma_i(j)=j, \,\,j\ne i,i+1,\label{sigmai}\ee
which satisfy the relations 
\be\sigma_i^2=1, \qquad (\sigma_i\sigma_{i+1})^3=1,\qquad (\sigma_i\sigma_j)^2=1,\,\,
|i-j|>1.\label{sigmarel}\ee
Let $\tilde \fS_n$ be the free group on the generators  $\{\tilde\sigma_i : 1\leq i\leq n-1\}$  and $\fW_n$ the smallest normal subgroup of $\tilde \fS_n$ containing  $\{\tilde \sigma_i^2,\, 1\leq i\leq n-1; \quad (\tilde\sigma_i\tilde\sigma_{i+1})^3,\, 1
\leq i\leq n-2;\quad (\tilde\sigma_i\tilde\sigma_j)^2\,|i-j|>1\}$. Then $\tilde \fS_n/\fW_n\cong \fS_n$ with 
$\tilde \sigma_i\mapsto \sigma_i$ defining an homomorphism $\tilde \fS_n\rightarrow \fS_n$, which we 
shall denote by $\tilde \varrho\mapsto \varrho$. (See \cite{CM}  page 63.) Each $\tilde\varrho\in\tilde \fS_n$ can be written 
$\varrho=\tilde\sigma_{i_1}\tilde\sigma_{i_2}\ldots\tilde\sigma_{i_k}$, where
$1\leq |i_j|\leq n-1$ and $\sigma_i^{-1}=\sigma_{-i}$. We can define a function 
$\tilde\phi : \tilde \fS_n\rightarrow \K_n$ in terms of $\tilde\phi(\tilde\sigma_i)$, $1\leq i\leq n-1$, 
with $\tilde\phi(\tilde\sigma_i^{-1})=\tilde\phi(\tilde\sigma_i)$ and $\tilde\phi(1)=0$, by 
\be\tilde\phi(\tilde\sigma_{i_1}\ldots\tilde\sigma_{i_k})=\sum_{j=1}^k\sigma_{i_1}\ldots\sigma_{i_{j-1}}\tilde\phi(\tilde\sigma_{i_j}).\label{gendeftphi}\ee
Then 
\be\tilde\phi(\tilde\varrho_1\tilde\varrho_2)=\varrho_1\tilde\phi(\tilde\varrho_2)+\tilde\phi(\tilde\varrho_1).
\label{phirhorho}\ee

We now show that $\ker \tilde\phi\cap \fW_n$ is a normal subgroup of $\tilde\fS_n$. Suppose 
$\tilde\varrho\in\ker \tilde\phi\cap \fW_n$, so that $\varrho=1\in\fS_n$ and $\tilde\phi(\tilde\varrho)=0$. Then
$$\tilde\phi(\tilde\varrho_1\tilde\varrho\varrho_1^{-1})=\varrho_1\varrho\tilde\phi(\tilde\varrho_1^{-1})+\varrho_1\tilde\phi(\tilde\varrho)
+\tilde\phi(\tilde\varrho_1)=\varrho_1\tilde\phi(\tilde\varrho_1^{-1})
+\tilde\phi(\tilde\varrho_1)=\tilde\phi(1)=0$$
so that $\tilde\varrho_1\tilde\varrho\tilde\varrho_1^{-1}\in \ker \tilde\phi\cap \fW_n$ and this is a normal subgroup of $\tilde \fS_n$. So if we can show that
\be\{\tilde \sigma_i^2,\, 1\leq i\leq n-1; \, (\tilde\sigma_i\tilde\sigma_{i+1})^3,\, 1
\leq i\leq n-2;\, (\tilde\sigma_i\tilde\sigma_j)^2,\,|i-j|>1\}\subset\ker\tilde\phi\label{kerphi}\ee
we must have $\ker \tilde\phi\cap \fW_n=\fW_n$, {\it i.e.}  $\fW_n\subset\ker\tilde\phi$, because
$\fW_n$ is the smallest normal subgroup containing these elements.
Then $\tilde\phi$ induces a map $\phi:\fS_n\rightarrow \K_n$ with $\phi(\varrho)=\tilde\phi(\tilde\varrho)$, 
because $\phi(\tilde\varrho\tilde w)=\phi(\tilde\varrho)+\varrho\phi(\tilde w)=\phi(\tilde \varrho)$ if $\tilde w\in \fW_n$.

Next we show that \eqref{kerphi} holds if we define
\be\tilde\phi(\tilde\sigma_i,\kappa_{n-1})={f^{\alpha_i\alpha_{i+1}}}_\beta  \kappa_{n-1}^{\alpha_1\ldots\alpha_{i-1}\beta\alpha_{i+2}\ldots\alpha_n}\in\T_n,\label{deftildephi}\ee
for  $\kappa_{n-1}\in\K_{n-1}$.
We will write out the argument for $\tilde\sigma_1^2, (\tilde\sigma_1\tilde\sigma_2)^3, (\tilde\sigma_1\tilde\sigma_3)^2$ and the arguments for other values of $i, j$ follow by similar arguments. First,
writing $\kappa=\kappa_{n-1}$,
$$
\tilde\phi(\tilde\sigma_1^2,\kappa)=\tilde\phi(\tilde\sigma_1,\kappa)+\sigma_1\tilde\phi(\tilde\sigma_1,\kappa),
$$
implying
$$
\tilde\phi(\tilde\sigma_1^2,\kappa)^{\alpha_1\alpha_2\alpha_3\ldots\alpha_n}
={f^{\alpha_1\alpha_2}}_\beta\kappa^{\beta\alpha_3\ldots\alpha_n}
+{f^{\alpha_2\alpha_1}}_\beta\kappa^{\beta\alpha_3\ldots\alpha_n}=0.
$$
Second
\begin{align}
\tilde\phi((\tilde\sigma_1\tilde\sigma_3)^2,\kappa)&=\tilde\phi(\tilde\sigma_1,\kappa)+\sigma_1\tilde\phi(\tilde\sigma_3,\kappa)+\sigma_1\sigma_3\tilde\phi(\tilde\sigma_1,\kappa)+\sigma_1\sigma_3\sigma_1\tilde\phi(\tilde\sigma_3,\kappa)\cr
&=\tilde\phi(\tilde\sigma_1,\kappa)+\sigma_1\sigma_3\tilde\phi(\tilde\sigma_1,\kappa)+\sigma_1\tilde\phi(\tilde\sigma_3,\kappa)+\sigma_3\tilde\phi(\tilde\sigma_3,\kappa)
\end{align}
implying
\begin{align}
\tilde\phi((\tilde\sigma_1\tilde\sigma_3)^2,\kappa)^{\alpha_1\alpha_2\alpha_3\alpha_4\ldots\alpha_n}
&={f^{\alpha_1\alpha_2}}_\beta\kappa^{\beta\alpha_3\alpha_4\ldots\alpha_n}
+{f^{\alpha_2\alpha_1}}_\beta\kappa^{\beta\alpha_4\alpha_3\ldots\alpha_n}\cr
&\hskip20truemm+{f^{\alpha_3\alpha_4}}_\gamma\kappa^{\alpha_2\alpha_1\gamma\ldots\alpha_n}
+{f^{\alpha_4\alpha_3}}_\gamma\kappa^{\alpha_1\alpha_2\gamma\ldots\alpha_n}\cr
&={f^{\alpha_1\alpha_2}}_\beta{f^{\alpha_3\alpha_4}}_\gamma\kappa^{\beta\gamma\alpha_5\ldots\alpha_n}
+{f^{\alpha_2\alpha_1}}_\beta{f^{\alpha_3\alpha_4}}_\gamma\kappa^{\beta\gamma\alpha_5\ldots\alpha_n}
=0.\end{align}
Third,
\begin{align}
\tilde\phi((\tilde\sigma_1&\tilde\sigma_2)^3,\kappa)\cr
&=\tilde\phi(\tilde\sigma_1,\kappa)+\sigma_1\tilde\phi(\tilde\sigma_2,\kappa)+\sigma_1\sigma_2\tilde\phi(\tilde\sigma_1,\kappa)+\sigma_1\sigma_2\sigma_1\tilde\phi(\tilde\sigma_2,\kappa)+\sigma_2\sigma_1\tilde\phi(\tilde\sigma_1,\kappa)+\sigma_2\tilde\phi(\tilde\sigma_2,\kappa)\cr
&={f^{\alpha_1\alpha_2}}_\beta\kappa^{\beta\alpha_3\alpha_4\ldots\alpha_n}+
{f^{\alpha_1\alpha_3}}_\beta\kappa^{\alpha_2\beta\alpha_4\ldots\alpha_n}+
{f^{\alpha_2\alpha_3}}_\beta\kappa^{\beta\alpha_1\alpha_4\ldots\alpha_n}+
{f^{\alpha_2\alpha_1}}_\beta\kappa^{\alpha_3\beta\alpha_4\ldots\alpha_n}\cr
&\hskip2truecm+
{f^{\alpha_3\alpha_1}}_\beta\kappa^{\beta\alpha_2\alpha_4\ldots\alpha_n}+
{f^{\alpha_3\alpha_2}}_\beta\kappa^{\alpha_1\beta\alpha_4\ldots\alpha_n}\cr
&={f^{\alpha_1\alpha_2}}_\beta {f^{\beta\alpha_3}}_\gamma \kappa^{\gamma\alpha_4\ldots\alpha_n}+
{f^{\alpha_1\alpha_3}}_\beta {f^{\alpha_2\beta}}_\gamma \kappa^{\gamma\alpha_4\ldots\alpha_n}+
{f^{\alpha_2\alpha_3}}_\beta {f^{\beta\alpha_1}}_\gamma \kappa^{\gamma\alpha_4\ldots\alpha_n}
=0\end{align}
by the Jacobi identity. This establishes \eqref{kerphi}.

Thus, for each $\kappa_{n-1}\in\K_{n-1}$, we can define 
$\phi(\varrho,\kappa_{n-1})=\tilde\phi(\tilde\varrho,\kappa_{n-1})$ 
and define $\phi:\K_{n-1}\rightarrow\K_n$ by
\be\phi(\kappa_{n-1})={1\over n!}\sum_{\varrho\in \fS_n}\phi(\varrho,\kappa_{n-1}).\label{phidef}\ee
We shall now show that $\phi(\kappa_{n-1})$ satisfies \eqref{kappaeq4}  and  \eqref{kcycle4}.
From \eqref{phirhorho},
\be\phi(\varrho_1\varrho_2,\kappa_{n-1})=\varrho_1\phi(\varrho_2,\kappa_{n-1})+\phi(\varrho_1,\kappa_{n-1}).
\label{phirhorho2}\ee

Then, for any $\sigma\in \fS_n$,
\begin{align}
\phi(\kappa_{n-1})-\sigma\phi(\kappa_{n-1})&={1\over n!}\sum_{\varrho\in \fS_n}\phi(\varrho,\kappa_{n-1})-{1\over n!}\sum_{\varrho\in \fS_n}\sigma\phi(\varrho,\kappa_{n-1})\cr
&={1\over n!}\sum_{\varrho\in \fS_n}\phi(\varrho,\kappa_{n-1})-{1\over n!}\sum_{\varrho\in \fS_n}\phi(\sigma\varrho,\kappa_{n-1})
+{1\over n!}\sum_{\varrho\in \fS_n}\phi(\sigma,\kappa_{n-1})\cr
&=\phi(\sigma,\kappa_{n-1})
\label{phisigma}
\end{align}
Taking $\sigma=\sigma_1$ in \eqref{phisigma}, we have
$$
\phi(\kappa_{n-1})^{\alpha_1\alpha_2\alpha_3\ldots\alpha_n}
-\phi(\kappa_{n-1})^{\alpha_2\alpha_1\alpha_3\ldots\alpha_n}={f^{\alpha_1\alpha_2}}_\beta  \kappa_{n-1}^{\beta\alpha_3\ldots\alpha_n}
$$
so that \eqref{kappaeq4} holds. 

If $\sigma=\sigma_1\sigma_2\ldots\sigma_{n-1}$, then $\sigma(j)=j+1$, $1\leq j\leq n-1$ and
$\sigma(n)=1$, {\it i.e.} $\sigma$ is cyclic permutation of $(1,2,\ldots,n)$
so that
\begin{align}
\phi(\sigma,\kappa_{n-1})&=\sum_{j=1}^{n-1}\sigma_{1}\ldots\sigma_{{j-1}}\phi(\sigma_{j},\kappa_{n-1})\cr
&=\sum_{j=1}^{n-1}\sigma_{1}\ldots\sigma_{{j-1}}{f^{\alpha_j\alpha_{j+1}}}_\beta
\kappa_{n-1}^{\alpha_1\ldots\alpha_{j-1}\beta\alpha_{j+2}\ldots\alpha_n}\cr
&=\sum_{j=1}^{n-1}{f^{\alpha_1\alpha_{j+1}}}_\beta
\kappa_{n-1}^{\alpha_2\ldots\alpha_{j}\beta\alpha_{j+2}\ldots\alpha_n}=0
\end{align}
as $\kappa_{n-1}\in\K_{n-1}$ is invariant. Thus putting $\sigma=\sigma_1\sigma_2\ldots\sigma_{n-1}$ in \eqref{phisigma} gives
$$\phi(\kappa_{n-1})^{\alpha_1\alpha_2\ldots\alpha_n}=\phi(\kappa_{n-1})^{\alpha_n\alpha_1\ldots\alpha_{n-1}},
$$
so that  \eqref{kcycle4} holds. 

By averaging \eqref{phisigma} over $\sigma\in\fS_n$, we see that
$$\sum_{\sigma\in\fS_n}\sigma\phi(\kappa_{n-1})=0,$$
so that it is orthogonal to all symmetric tensors and so the unique solution to \eqref{kappaeq4} and  \eqref{kcycle4} with this property.

This completes the proof of Proposition 2.

Note that Proposition 2 implies that $\K_n/\Sc_n\cong \K_{n-1}$, where $\Sc_n\subset \K_n$ is the space of symmetric invariant tensors.

As particular instances, we have that if $\kappa^{ab}$ is an invariant symmetric tensor, the general
solution for \eqref{kappaeq4} and  \eqref{kcycle4} for $n=3$ is 
\be\kappa_3^{abc} =\half {f^{ab}}_e\kappa^{ec}+\omega^{abc},\label{kappa3}\ee
where $\omega^{abc}$ is symmetric, and invariant for $\kappa_3$ to be invariant. In this case, the general solution to \eqref{kappaeq4} and  \eqref{kcycle4} for $n=4$ is
\be\kappa_4^{abcd} =\sixth  {f^{ab}}_e {f^{cd}}_g\kappa^{eg}+\sixth {f^{da}}_e {f^{bc}}_g\kappa^{eg}+
\half {f^{ab}}_e\omega^{ecd}+\half {f^{bc}}_e\omega^{ead}+
\half {f^{ac}}_e\omega^{ebd}+\omega^{abcd},\label{kappa4}\ee
where $\omega^{abcd}$ is symmetric.

While Proposition 2 proves the existence of a solution to \eqref{kappaeq4} and  \eqref{kcycle4}  for a given $\kappa_{n-1}\in\K_{n-1}$, it does not provide an explicit expression for such a solution unless we have a method of specifying expressions for each element $\varrho\in\fS_n$ as a product $\sigma_{i_1}\sigma_{i_2}\ldots\sigma_{i_k}$ of transpositions. In Appendix \ref{sect:constrkappa}, we derive an explicit expression for $\kappa_n$ in terms of $\kappa_{n-1}$ using the representation theory of $\fS_n$ and Young tableaux.
\vfil\eject
\section{Current Algebra on the Torus}
\label{sect:Loop}
\subsection{\sl  Fermionic Loop Constructions}
\label{fermionloop}

We consider the loop amplitude
\be\A_\floop^{a_1a_2\ldots a_n}(\nu_1,\nu_2,\ldots,\nu_n,\tau)
=\tr\left(J^{a_1}(\rho_1)J^{a_2}(\rho_2)\ldots J^{a_n}(\rho_n)w^{L_0}\right)
(2\pi i)^n\prod_{j=1}^n\rho_j,\label{Aloop}\ee
where
\be\rho_j=e^{2\pi i\nu_j},\qquad w=e^{2\pi i\tau}.\label{rhow}\ee
 and begin by reviewing the explicit expressions for this amplitude 
 when $J^a(\rho)$ is given as a bilinear in fermionic fields.
First, we take $J^a(\rho)$ to be given in terms of Neveu-Schwarz fields by \eqref{Mbb}.
Defining the partition function
\be\chi_{NS}(\tau)=\prod_{r=\half}^\infty (1+w^r)^N,\label{chiNS}\ee
we can write
\be
tr(b^i(\rho_1)b^j(\rho_2)w^{L_0})={1\over 2\pi i(\rho_1\rho_2)^\half}\,\chi_{NS}(\nu_1-\nu_2,\tau)\,\chi_{NS}(\tau)\delta^{ij},
\label{NSb2pt}
\ee
where
\be
\chi_{NS}(\nu,\tau)=2\pi i\sum_{r=\half}^\infty {e^{-2\pi i r\nu}+w^re^{2\pi i r\nu}\over 1+w^r}
={\theta_1'(0,\tau)\theta_3(\nu,\tau)\over\theta_3(0,\tau)\theta_1(\nu,\tau)}
\sim {1\over \nu}\quad\hbox{as}\quad \nu\rightarrow 0.\label{lab33}\ee

With $J^a(\rho)$ given by \eqref{Mbb}, the two-point function is
\begin{align}
\tr\left( J^a(\rho_1)J^b(\rho_2)w^{L_0}\right)
&=-{\kappa^{ab}\over4\pi^2\rho_1\rho_2}\chi_{NS}
(\nu_1-\nu_2,\tau)^2\chi_{NS}(\tau)\cr
&=-{\kappa^{ab}\over4\pi^2\rho_1\rho_2}\P_{NS}(\nu_1-\nu_2,\tau)
\chi_{NS}(\tau)\label{lab35}\end{align}
where $\kappa^{ab}=-\half\tr(M^aM^b)=\half\tr(t^at^b)$, and 
\begin{align}
\P_{NS}(\nu,\tau)&={\theta_1'(0,\tau)^2\theta_3(\nu,\tau)^2\over\theta_3(0,\tau)^2\theta_1(\nu,\tau)^2}
\sim{1\over\nu^2},\quad\hbox{as}\quad \nu\rightarrow 0\cr
&={\theta_3''(0,\tau)\over\theta_3(0,\tau)}-\left({\theta_1'(\nu,\tau)\over\theta_1(\nu,\tau)}\right)'\cr
&={\theta_3''(0,\tau)\over\theta_3(0,\tau)}+2\eta(\tau)+\P(\nu,\tau).\label{lab38}
\end{align}
Here the Weierstrass  $\P$ function,
\be\P(\nu,\tau)=-\left({\theta_1'(\nu,\tau)\over\theta_1(\nu,\tau)}\right)'-2\eta(\tau),\label{lab39}\ee
with
\be\eta(\tau)=-{1\over 6}{\theta'''_1(0,\tau)\over \theta_1'(0,\tau)}.\label{etatau}\ee
(See \cite{HTF}, page 361.)

The general prescription for the $n$-point loop amplitude \eqref{Aloop}, with $J^a(\rho)$ given by \eqref{Mbb}, is given by 
a modification of the Frenkel-Zhu construction of \S\ref{sect:FZ}, by writing \eqref{Aloop} as a sum over permutations $\rho\in\fS_n$ with no fixed point. If $\rho=\xi_1\xi_2\ldots\xi_r$, a product of disjoint cycles, we associate to $\rho$ a product
\be F_\rho^{NS}=(-1)^r f_{\xi_1}^{NS}f_{\xi_2}^{NS}\ldots f_{\xi_r}^{NS}\chi_{NS}(\tau),\label{Frho}\ee
 where the function $f_\xi^{NS}$ associated with the cycle $\xi=(i_1,i_2\ldots i_m)$ is defined by 
\be
f_\xi^{NS}=\kappa^{a_{i_1}a_{i_2}\ldots a_{i_m}}\chi_{NS}(\nu_{i_1}-\nu_{i_2},\tau)\chi_{NS}(\nu_{i_2}-\nu_{i_3},\tau)\ldots \chi_{NS}(\nu_{i_m}-\nu_{i_1},\tau)\label{fxi}
\ee
\be\kappa^{a_1a_2\ldots a_n}=\half\tr(t^{a_1}t^{a_2}\ldots t^{a_n})
=\half i^n\tr(M^{a_1}M^{a_2}\ldots M^{a_n}), \label{lab36}\ee
The $n$-point loop amplitude is then constructed as the sum of these products over the permutations $\rho\in \fS'_n$, the subset of $\fS_n$ with no fixed points,
\be 
\A_\floop^{a_1a_2\ldots a_n}(\nu_1,\nu_2,\ldots,\nu_n,\tau)
=\sum_{\rho\in\fS'_n} F^{NS\hskip1pt a_1a_2\ldots a_n}_\rho(\nu_1,\nu_2,\ldots,\nu_n,\tau).
\label{AloopNS}\ee
Again this construction can be described  graphically  by summing over all graphs with $n$ vertices where the  vertices carry the labels $1,2,\ldots, n.$ and each vertex is connected by directed lines to other vertices, one of the lines at each vertex pointing towards it and one away from it. An expression \eqref{fxi} is associated with each cycle, together with  factor of $-1$, and the product of these cycle expressions is associated with the whole graph.

For example, this gives as the expression for the three-point loop
\begin{align}
\tr\left(J^{a}(\rho_1)J^{b}(\rho_2)J^{c}(\rho_3)w^{L_0}\right)&\cr
&\hskip-20truemm={ -ik\over 8\pi^3 \rho_1\rho_2\rho_3}f^{abc}\chi_{NS}(\tau)
\chi_{NS}(\nu_1-\nu_2,\tau)\chi_{NS}(\nu_2-\nu_3,\tau)\chi_{NS}(\nu_3-\nu_1,\tau)\cr
\end{align}
if $\tr(M^aM^b)=-2k\delta^{ab}$, so that $\kappa^{ab}=k\delta^{ab}$, and $\delta^{ab}$ is used to raise and lower indices.

We can modify the above to give a second fermionic construction by defining the partition function
\be\chi_{NS}^-(\tau)=\prod_{r=\half}^\infty (1-w^r)^N,\label{chiNSm}\ee
we can write
\be
\tr\left(b^i(\rho_1)b^j(\rho_2)w^{L_0}(-1)^{N_b}\right)={1\over 2\pi i(\rho_1\rho_2)^\half}\chi_{NS}^-(\nu_1-\nu_2,\tau)\chi_{NS}^-(\tau)\delta^{ij},\label{bminus}
\ee
where
\be
\chi_{NS}^-(\nu,\tau)=2\pi i\sum_{r=\half}^\infty {e^{-2\pi i r\nu}-w^re^{2\pi i r\nu}\over 1-w^r}
={\theta_1'(0,\tau)\theta_4(\nu,\tau)\over\theta_4(0,\tau)\theta_1(\nu,\tau)}
\sim {1\over \nu}\quad\hbox{as}\quad \nu\rightarrow 0;\label{lab45}\ee
and, if we replace $\chi_{NS}(\nu,\tau)$ by $\chi_{NS}^-(\nu,\tau)$ in \eqref{fxi}, with $\chi_{NS}^-(\tau)$ replacing $\chi_{NS}(\tau)$ in \eqref{Frho}, the above construction for the loop amplitudes 
gives
\be\tr\left(J^{a_1}(\rho_1)J^{a_2}(\rho_2)\ldots J^{a_n}(\rho_n)w^{L_0}(-1)^{N_b}\right),\label{lab46}\ee
where $N_b=\sum_{r>0}b_{-r}b_r$.
In particular, the two-point function is
\begin{align}
\tr\left( J^a(\rho_1)J^b(\rho_2)w^{L_0}(-1)^{N_b}\right)
&=-{\kappa^{ab}\over4\pi^2\rho_1\rho_2}\chi_{NS}^-(\nu_1-\nu_2,\tau)^2\chi_{NS}^-(\tau)\cr
&=-{\kappa^{ab}\over4\pi^2\rho_1\rho_2}\P_{NS}^-(\nu_1-\nu_2,\tau)\chi_{NS}^-(\tau)
\end{align}
where
\begin{align}
\P_{NS}^-(\nu,\tau)&={\theta_1'(0,\tau)^2\theta_4(\nu,\tau)^2\over\theta_4(0,\tau)^2\theta_1(\nu,\tau)^2}\cr
&={\theta_4''(0,\tau)\over\theta_4(0,\tau)}-\left({\theta_1'(\nu,\tau)\over\theta_1(\nu,\tau)}\right)'\cr
&={\theta_4''(0,\tau)\over\theta_4(0,\tau)}+2\eta(\tau)+\P(\nu,\tau).
\end{align}

A third fermionic construction is given by using the Ramond operators,
\be d^i(\rho)=\sum_{m\in\Zop}d^i_m\rho^{-m+\half},\qquad \{d_m^i,d_n^j\}=\delta_{m,-n}\delta^{ij},\qquad
d_m^i\vac=0, \quad m>0,\qquad 1\leq i,j,\leq N.\label{dalg}\ee
Defining the Ramond partition function
 \be\chi_R(\tau)=\prod_{n=1}^\infty (1+w^n)^N,\label{chiR}\ee
 we can write
$$
\tr\left(d^i(\rho_1)d^j(\rho_2)w^{L_0}\right)={1\over 2\pi i(\rho_1\rho_2)^\half}\\chi_R(\nu_1-\nu_2,\tau)\chi_R(\tau)\delta^{ij},
$$
where
\be
\chi_R(\nu,\tau)=\pi i+2\pi i \sum_{m=1}^\infty {e^{-2\pi i m\nu}+w^me^{2\pi i m\nu}\over 1+w^m}
={\theta_1'(0,\tau)\theta_2(\nu,\tau)\over\theta_2(0,\tau)\theta_1(\nu,\tau)}
\sim {1\over \nu}\quad\hbox{as}\quad \nu\rightarrow 0.\label{lab51}
\ee
If we now replace $\chi_{NS}(\nu,\tau)$ by $\chi_{R}(\nu,\tau)$ in \eqref{fxi}, with $\chi_{R}(\tau)$ replacing $\chi_{NS}(\tau)$ in \eqref{Frho}, the construction for the loop amplitudes 
gives
\be\tr\left(J^{a_1}(\rho_1)J^{a_2}(\rho_2)\ldots J^{a_n}(\rho_n)w^{L_0}\right),\label{lab52a}\ee
where now, instead of \eqref{Mbb},
\be J^a(\rho)={i\over 2}M^a_{ij}d^i(\rho)d^j(\rho).\label{lab52b}\ee
The two-point function is now
\begin{align}
\tr\left( J^a(\rho_1)J^b(\rho_2)w^{L_0}\right)
&=-{\kappa^{ab}\over4\pi^2\rho_1\rho_2}\chi_{R}(\nu_1-\nu_2,\tau)^2\chi_{R}(\tau)\cr
&=-{\kappa^{ab}\over4\pi^2\rho_1\rho_2}\P_{R}(\nu_1-\nu_2,\tau)\chi_{R}(\tau)
\end{align}
where
\begin{align}
\P_R(\nu,\tau)&={\theta_1'(0,\tau)^2\theta_2(\nu,\tau)^2\over\theta_2(0,\tau)\theta_1(\nu,\tau)^2}\cr
&={\theta_2''(0,\tau)\over\theta_2(0,\tau)}-\left({\theta_1'(\nu,\tau)\over\theta_1(\nu,\tau)}\right)'\cr
&={\theta_2''(0,\tau)\over\theta_2(0,\tau)}+2\eta(\tau)+\P(\nu,\tau).
\end{align}

\subsection{\sl  General Torus Amplitudes and Connected Parts}
\label{connected}

The loop amplitude
\be\A_\floop^{a_1a_2\ldots a_n}(\nu_1,\nu_2,\ldots,\nu_n,\tau)
=\tr\left(J^{a_1}(\rho_1)J^{a_2}(\rho_2)\ldots J^{a_n}(\rho_n)w^{L_0}\right)
(2\pi i)^n\prod_{j=1}^n\rho_j,\label{Aloop2}\ee
 is invariant under $\nu_j\rightarrow\nu_j+1$ and $\nu_j\rightarrow\nu_j+\tau$ for each $j$ individually, so that it is defined on the torus obtained by identifying $\nu\in\Cop$ with $\nu+1$ and $\nu+\tau$. Because of the locality of the currents $J^{a_j}(\rho_j)$, the amplitude is also symmetric under simultaneous permutations of the $\rho_j$ and the $a_j$.

From \eqref{ope} we have that
\be
J^a(\rho_1)J^b(\rho_2)\sim
{\kappa^{ab}\over(2\pi i)^2\rho_1 \rho_2
(\nu_1-\nu_2)^2}+  {{f^{ab}}_cJ^c(\rho_2)\over2\pi i \rho_1
(\nu_1-\nu_2)}\quad\hbox{as}\quad \nu_1\rightarrow\nu_2.\label{ope2}\ee
Thus the singularities of the $n$-point loop amplitude on the torus are determined in terms of the $(n-1)$-point and $(n-2)$-point loop amplitudes. This means that knowledge of the $(n-1)$-point and $(n-2)$-point loop amplitudes determines the $n$-point loop amplitude up to a constant on the torus, that is a function of $\tau$.  (See, {\it e.g.},  \cite{RM}, page 29.)  Because of the permutation symmetry of the amplitude \eqref{Aloop2}, this leaves the $n$-point loop determined up to a symmetric invariant tensor function of $\tau$, given the $(n-1)$-point and $(n-2)$-point loops.

The sum over permutations in the expression \eqref{AloopNS} for the loop in the fermionic construction cases can be divided into terms which collect together the same $\rho_i$ in each cycle. Such terms are labeled by the division of the variables $\{\rho_1,\rho_2,\ldots, \rho_n\}$ into subsets, each consisting of at least two elements (corresponding to the restriction to permutations with no fixed points). The full loop amplitude is then the sum over these terms. Such terms are products of ``connected parts'', each of which involves one of the subsets of $\{\rho_1,\rho_2,\ldots, \rho_n\}$, say $\{\rho_{i_1},\rho_{i_2},\ldots, \rho_{i_m}\}$, given by an expression like \eqref{Atreef1}, 
\be\A_{\floop, C}^{a_1a_2\ldots a_m}(\nu_1,\nu_2,\ldots, \nu_m,\tau)
=-{1\over m}\sum_{\varrho\in \fS_m}f^{NS}_{(\varrho(1),\varrho(2),\ldots,\varrho(m))}\chi_{NS}(\tau),\label{lab55}\ee
in the NS case. The amplitudes $\A_{\floop, C}$ have a simpler structure 
than the full amplitudes
$\A_\floop$ in that they have only single poles for $m>2$, 
rather than both single and double poles.
For $m>2$, the connected amplitudes satisfy the conditions
\be\A_{\floop, C}^{a_1a_2\ldots a_m}(\nu_1,\nu_2,\ldots, \nu_m,\tau)
\sim {1\over \nu_m-\nu_j}
{f^{a_ma_j}}_{a_j'}\A_{\floop, C}^{a_1a_2\ldots a_{j-1}a_j'a_{j+1}\ldots a_{m-1}}(\nu_1,\nu_2,\ldots, \nu_{m-1},\tau),\label{Aloopres}\ee
which are sufficient to specify the $m$-point connected amplitude $\A_{\floop, C}$ in terms of the
$(m-1)$-point connected amplitude, again up to a symmetric invariant tensor function of $\tau$.

Motivated by the fermionic constructions, we can give a general definition of the connected part of the loop amplitude in a familiar way. If $A=\{i_1,i_2,\ldots, i_n\}$ is a set of distinct positive integers, define
\be\A^A\equiv \A_{ \rm loop}^{a_{i_1}a_{i_2}\ldots a_{i_n}}(\nu_{i_1},\nu_{i_2},\ldots,\nu_{i_n},\tau)
=\tr\left(J^{a_{i_1}}(\rho_{i_1})J^{a_{i_2}}(\rho_{i_2})\ldots J^{a_{i_n}}(\rho_{i_n})w^{L_0}\right)
(2\pi i)^n\prod_{j=1}^n\rho_{i_j}.\label{AA}\ee
Let $P=(A_1,A_2,\ldots, A_r)$ be a division of the integers $A=\{i_1,i_2,\ldots, i_n\}=A_1\cup A_2\cup\ldots\cup A_r$ into a number of disjoint subsets; let $\fP_A$ denote the collections of such divisions; and denote the partition function by
\be\chi(\tau)=\tr\left(w^{L_0}\right)\label{partfn}.\ee
Then we can define the connected amplitude $\A^A_C$ inductively by 
\be\A^A=\sum_{P\in\fP_A}\chi(\tau)^{1-|P|}\prod_{A_j\in P}\A^{A_j}_C,\label{induct}\ee
where $|P|=r$, the number of subsets contained in the division $P$,
together with the vanishing of the one point function $\A^{\{i\}}_C=0$, and, consequently,
\be\A^{\{i,j\}}_C=\A^{\{i,j\}}=\tr\left(J^{a_{i}}(\rho_{i})J^{a_{j}}(\rho_{j})w^{L_0}\right)
(2\pi i)^2\rho_{i}\rho_{j}.\label{AijC}\ee

Equation \eqref{induct}  is of the form given for the NS case by \eqref{AloopNS} together with \eqref{Frho}, where
$$\A^{\{i_1\ldots i_m\}}_C=-\sum_{\varrho\in\fS_m}f^{NS}_{i_{\varrho(1)}\ldots i_{\varrho(m)}}\chi,
\qquad \chi(\tau)=\chi_{NS}(\tau).$$

Equation \eqref{induct} defines an inductive procedure because we can write it as
\be\A^A_C=\A^A-\sum_{P\in\fP_A'}\chi(\tau)^{1-|P|}\prod_{A_j\in P}\A^{A_j}_C,\label{inductproc}\ee
where $\fP_A'$ denotes the same collection of divisions of $A$ into disjoint subsets but omitting the division of $A$ into the single set consisting of itself.  If we single out a point $i\in A$, we can rewrite the inductive definition of $\A^A_C$,
\be\A^A=\A^A_C+\sum_{B\in\fR_A^i} \A_C^B\A^{A\sim B}/\chi,\label{AAAC}\ee
where $\fR_A^i$ denotes the proper subsets of $\A$ which contain $i$.

The point of this definition of 
\be\A^A_C\equiv\A^{a_{i_1}a_{i_2}\ldots a_{i_n}}_C(\nu_{i_1},\nu_{i_2},\ldots,\nu_{i_n},\tau)\label{lab121}\ee
is that, for $m>2$, the double poles at $\nu_i=\nu_j$ present in $\A^A$ have been removed and only single poles remain. A double pole remains in $\A^{\{i,j\}}_C$ defined by \eqref{AijC},
$$\A^{\{i,j\}}_C\sim{\kappa^{a_ia_j}\over(\nu_i-\nu_j)^2}\chi(\tau)\qquad\hbox{as }\nu_i\sim\nu_j,$$
and this is its only singularity.
 To demonstrate the absence of the double pole at $\nu_i=\nu_j$ in $\A^A_C$, $m>2$, we use induction and \eqref{AAAC}. We note that the residue of the double pole on the left hand side is $k\delta^{a_ia_j}\A^{A\sim{\{i,j\}}}$ and in the sum on the right hand side, assuming inductively that the result is true for smaller amplitudes, the double pole occurs only in the term involving $\A^B_C$ for $B={\{i,j\}}$ and the residue for this term is $\delta^{a_ia_j}k\chi$ multiplied by $\A^{A\sim B}/\chi$, {\it i.e.} the same as on the left hand side, so that these residues cancel and $\A_C^A$ has no double pole at $\nu_i=\nu_j$. A similar argument shows that $\A^A_C$ satisfies the same relations for the residues at single poles as $\A^A$, so that \eqref{Aloopres} holds.

\subsection{\sl  Structure of Torus Amplitudes}
\label{structuretorus}

 In general write
\be\A_{\floop, C}^{a_1a_2\ldots a_n}(\nu_1,\nu_2,\ldots, \nu_n,\tau)
\equiv-\F_n^{a_1a_2\ldots a_n}(\nu_1,\nu_2,\ldots, \nu_n,\tau)\chi(\tau),\label{defF}\ee

so that, for $n>2$,
\be\F_n^{a_1a_2\ldots a_n}(\nu_1,\nu_2,\ldots, \nu_n,\tau)
\sim {1\over \nu_n-\nu_j}
{f^{a_ma_j}}_{a_j'}\F_{n-1}^{a_1a_2\ldots a_{j-1}a_j'a_{j+1}\ldots a_{n-1}}(\nu_1,\nu_2,\ldots, \nu_{n-1},\tau)\label{Fres}\ee
as $\nu_n\sim\nu_j$, which specifies $\F_n$ on the torus in terms of $\F_{n-1}$ up to a function of $\tau$,
\be\omega_n^{a_1a_2\ldots a_n}(\tau),\label{lab57}\ee
which, because of the properties of $\F_n$, must be an invariant symmetric tensor. 
Inductively, this determines $\F_n$ in terms of $\F_2$ and these invariant tensors, $\omega_m$,
$2<m\leq n$.
The 2-point function, $\F_2$, has only a double pole,
\be\F_2^{ab}(\nu_1,\nu_2,\tau)\sim
-{\kappa^{ab}\over(\nu_1-\nu_2)^2}\quad\hbox{as}\quad \nu_1\rightarrow\nu_2,\label{Fres2}\ee
In general \eqref{Fres2} implies that the general form of the two-point function is
\be\F^{ab}(\nu_1,\nu_2,\tau)=-\kappa^{ab}\P(\nu_1-\nu_2,\tau)
+\omega_2^{ab}(\tau),\label{gen2pt}\ee
where $\omega_2^{ab}(\tau)$ is a symmetric invariant tensor. In the NS, $\hbox{NS}^-$ and R cases, 
\be\omega_2^{ab}(\tau)=-\kappa^{ab}\left[
{\theta_s''(0,\tau)\over\theta_s(0,\tau)}+2\eta(\tau)\right]\label{lab60}\ee
with  $s=3,4,2,$ respectively. 

We can construct the general three-point loop, $\F_3$, as follows; we start by rewriting 
\eqref{gen2pt} as
\be\F_2^{ab}(\nu_1,\nu_2,\tau)=-\kappa^{ab}\P_{NS}(\nu_1-\nu_2,\tau)
+\tilde\omega_2^{ab}(\tau).\label{lab61}\ee
We then have that $\F^{abc}(\nu_1,\nu_2,\nu_3,\tau)$ differs from what it is in the NS case,
\be kf^{abc}
\chi_{NS}(\nu_1-\nu_2,\tau)\chi_{NS}(\nu_2-\nu_3,\tau)\chi_{NS}(\nu_3-\nu_1,\tau),\label{lab62}\ee
by a function defined on torus, whose residues at $\nu_1=\nu_2$, $\nu_2=\nu_3$, $\nu_3=\nu_1$ are all $i{f^{ab}}_e\tilde\omega_2^{ec}(\tau)$.
To construct such a function, consider the Weierstrass $\zeta$ function (see \cite{WW}, page 445),
\be\zeta(\nu,\tau)={\theta_1'(\nu,\tau)\over\theta_1(\nu,\tau)}
+  2\eta(\tau)\nu  \label{lab63}\ee
$\zeta(\nu,\tau)$ has the properties:
\be\zeta(\nu+1,\tau)=\zeta(\nu,\tau)+2\eta(\tau),\qquad
\zeta(\nu+\tau,\tau)=\zeta(\nu,\tau)+2\eta(\tau)\tau-2\pi i,\label{lab64}\ee
\be\zeta'(\nu,\tau)=-\P(\nu,\tau),\qquad\zeta(-\nu,\tau)=-\zeta(\nu,\tau),\qquad
\zeta(\nu,\tau)={1\over\nu}+{\cal O}(\nu^3),\quad\hbox{as}\quad\nu\rightarrow 0.\label{lab65}\ee
It follows that 
\be\zeta(\nu_1-\nu_2,\tau)+\zeta(\nu_2-\nu_3,\tau)+\zeta(\nu_3-\nu_1,\tau)\label{lab66}\ee
is defined on the torus and has residue $1$ at  $\nu_1=\nu_2$, $\nu_2=\nu_3$ and $\nu_3=\nu_1$.
Thus the general form for 
\begin{align}
\F_3^{abc}(\nu_1,\nu_2,\nu_3,\tau)=k&f^{abc}
\chi_{NS}(\nu_1-\nu_2,\tau)\chi_{NS}(\nu_2-\nu_3,\tau)\chi_{NS}(\nu_3-\nu_1,\tau)\cr
&+f^{abe}\tilde\omega_2^{ec}(\tau)\left[\zeta(\nu_1-\nu_2,\tau)
+\zeta(\nu_2-\nu_3,\tau)+\zeta(\nu_3-\nu_1,\tau)\right]+ 2 
\omega_3^{abc}(\tau),\label{Fthree}
\cr\end{align}
where $\omega_3$ is a symmetric invariant tensor, because this has the residues specified by \eqref{Fres}. 
We could proceed to express the $n$-point connected loop amplitude as the expression in the NS case
\eqref{fxi} with additional terms, but, instead, we adopt an approach that is more symmetric between all the terms. To this end, we define functions, $H_{n,m}(\mu_1,\ldots,\mu_n,\tau)$, symmetric under the permutations of the $\mu_i$, initially for $0\leq m\leq 4$, by 
\begin{align}
H_{n,0}(\bmu,\tau)&=1,\cr
H_{n,1}(\bmu,\tau)&=\sum_{j=1}^n\zeta_j,\cr
2H_{n,2}(\bmu,\tau)&=\left(\sum_{j=1}^n\zeta_j\right)^2+\sum_{j=1}^n\zeta_j',\cr
6H_{n,3}(\bmu,\tau)&=\left(\sum_{j=1}^n\zeta_j\right)^3+
3\sum_{j=1}^n\zeta_j\sum_{j=1}^n\zeta_j'+\sum_{j=1}^n\zeta_j'',\cr
24H_{n,4}(\bmu,\tau)&=\left(\sum_{j=1}^n\zeta_j\right)^4+
6\left(\sum_{j=1}^n\zeta_j\right)^2\sum_{j=1}^n\zeta_j'
+4\sum_{j=1}^n\zeta_j\sum_{j=1}^n\zeta_j''+3\left(\sum_{j=1}^n\zeta_j'\right)^2
+\sum_{j=1}^n\zeta_j'''+k_4,\cr
\displaybreak[2]\label{explicitH}
\end{align}
where $\bmu=(\mu_1,\ldots,\mu_n)$, $\zeta_j=\zeta(\mu_j,\tau)$, and $k_4 = k_4(\tau)$ is a constant on the torus to be determined. Then
the singularities in the $\mu_j$ of
$H_{n,m}(\mu_1,\ldots,\mu_n,\tau)$ are simple poles at $\mu_j=0$ for $n>2$, and the residue
\be\Res{\mu_n}
H_{n,m}(\mu_1,\ldots,\mu_n,\tau)=
H_{n-1,m-1}(\mu_1,\ldots,\mu_{n-1},\tau),\label{lab69}\ee
for $1\leq m\leq 4$ and $n>2$.
This can be verified case by case but we shall give a general argument below.

The $H_{n,m}(\bmu,\tau)$ are not single valued for $\mu_j$ on the torus but, if we impose the constraint that $\mu_1+\ldots+\mu_n=0$, they are. So $H_{n,m}(\nu_{12},\ldots,\nu_{n1},\tau)$,
where $\nu_{ij}=\nu_i-\nu_j$ is defined on the torus and, for $n>2$, just has poles at $\nu_i=\nu_{i+1}$, $1\leq i\leq n$, with $\nu_{n+1}\equiv\nu_1$. For $n=2$, 
\be H_{2,1}(\nu_{12},\nu_{21},\tau)=0,\qquad H_{2,2}(\nu_{12},\nu_{21},\tau)=-\P(\nu_{12},\tau).\label{lab70}\ee

By Proposition 2 of section \ref{sect:Tensors}, we can define $n$-th order tensors $\kappa_{n,m}(\tau)$,
$n\geq m\geq 0$, $n\geq 2$,
by the conditions
\be\kappa_{n,m}^{a_1a_2\ldots a_n}(\tau)-\kappa_{n,m}^{a_2a_1\ldots a_n}(\tau)
={f^{a_1a_2}}_b\kappa_{n-1,m-1}^{ba_3\ldots a_n}(\tau),\label{kappatau1}\ee
\be\kappa_{n,m}^{a_1a_2\ldots a_n}(\tau)=\kappa_{n,m}^{a_2\ldots a_{n}a_{1}}(\tau),\label{kappatau2}\ee
together with the requirement that $\kappa_{n,m}$ be orthogonal to all symmetric tensors for $m>0$ and $n>2$, and the initial condition that $\kappa_{2,2}=\kappa, \kappa_{2,1}=0$ and $\kappa_{n,0}(\tau)=\omega_n(\tau)$, a symmetric invariant tensor. 
Then, setting 
\be\langle\kappa_{n,m}H_{n,m}\rangle^{a_1a_2\ldots a_n}(\nu_1,\nu_2,\ldots, \nu_n,\tau)=
{1\over n}\sum_{\varrho\in\fS_n}\kappa_{n,m}^{a_{\varrho(1)}\ldots a_{\varrho(n)}}H_{n,m}(\nu_{\varrho(1)\varrho(2)},\ldots,\nu_{\varrho(n)\varrho(1)},\tau),\label{avkH}\ee

$\F_n= \langle\kappa_{n,n-m}H_{n,n-m}\rangle$, $n\geq m$, provides a solution to 
\eqref{Fres} for each $m$. By the linearity of those equations, we obtain the solution,
\be \F_n=\sum_{m=0}^n\langle\kappa_{n,m}H_{n,m}\rangle,\qquad n\geq 2.\label{FkH}\ee
Because so far we only have $H_{n,m}$ for $0\leq m\leq 4$, \eqref{FkH} is only valid for $2\leq n\leq 4$.
Explicitly,
\begin{align}
\F_2&=\langle\kappa_2 H_{2,2}\rangle+\langle\kappa_{2,0}\rangle\cr
\F_3&=\langle\kappa_3 H_{3,3}\rangle+\langle\kappa_{3,1}H_{3,1}\rangle+\langle\kappa_{3,0}\rangle\cr
\F_4&=\langle\kappa_4 H_{4,4}\rangle+\langle\kappa_{4,2}H_{4,2}\rangle+\langle\kappa_{4,1}H_{4,1}\rangle+\langle\kappa_{4,0}\rangle,\label{Gscheme}
\end{align}
where we have written $\kappa_n\equiv\kappa_{n,n}$.

To demonstrate that $H_{n,m}$ has the desired properties, $0\leq m\leq 4$, and to extend its definition to higher values of $m$, we note we can write
\be H_{n,m}(\bmu,\tau)={1\over m!}\left[\sum_{j=1}^n(\partial_j+\zeta_j)\right]^m1,\qquad\hbox{for }1\leq m\leq 3,\label{Gnmform}\ee
where $\partial_j=\partial/\partial\mu_j$. (Here, and in what follows, $n\geq 2$.) The $\zeta$ function can be written in terms of the Weierstrass $\sigma$ function (see \cite{WW}, page 447)
\be \zeta(\mu,\tau)={\sigma'(\mu,\tau)\over\sigma(\mu,\tau)},\qquad \sigma(\mu,\tau)=e^{\eta(\tau)\mu^2}{\theta_1(\mu,\tau)\over
\theta'_1(0,\tau)},\label{lab83}\ee
with $\sigma(-\mu,\tau)=-\sigma(\mu,\tau)$, and
\begin{align}
\sigma(\mu,\tau)=\sum_{s=0}^\infty f_s(\tau)\mu^{2s+1}=\mu+ f_2(\tau)\mu^5+\ldots,
\qquad\hbox{because }f_1=0.
\label{sigmanu}
\end{align}
Then, defining $\hat H_{n,m}(\bmu,\tau)$ by the right hand side of  \eqref{Gnmform} for all $m\geq 0$, we have as $\mu_n\rightarrow 0$,
\begin{align}
\hat H_{n,m}(\bmu,\tau)&={1\over m!\sigma(\mu_n,\tau)}\left[\partial_n+\sum_{j=1}^{n-1}(\partial_j+\zeta_j)\right]^m
\sigma(\mu_n,\tau)\cr
&={1\over\mu_n m!}\left[\partial_n+\sum_{j=1}^{n-1}(\partial_j+\zeta_j)\right]^m
\sum_{s=0}^\infty f_s\mu_n^{2s+1}+{\cal O}(1)\cr
&={1\over\mu_n}\sum_{s=0}^{[\half m-\half]} f_s(\tau)\hat H_{n-1,m-2s-1}(\bmu',\tau)
+{\cal O}(1),
\end{align}
where $\bmu'=(\mu_1,\ldots,\mu_{n-1})$ and $[\half m-\half]$ is the greatest integer less than or equal to $\half m-\half$. 
Thus
\be\Res{\mu_n}
\hat H_{n,m}(\bmu,\tau)=
\sum_{s=0}^{[\half m-\half]}f_s(\tau)\hat H_{n-1,m-2s-1}(\bmu,\tau);\label{Gnmres}\ee
In particular,
$$
\Res{\mu_n}
\hat H_{n,m}(\bmu,\tau)=\hat H_{n-1,m-1}(\bmu',\tau),\qquad 1\leq m\leq 4,
$$
so that \eqref{lab69} holds for $1\leq m\leq 4$ and $n>2$, but
$$\Res{\mu_n}\hat H_{n,5}(\bmu,\tau)=\hat H_{n-1,4}(\bmu,\tau)+f_2\hat H_{n-1,0}(\bmu',\tau).$$

To see how to modify $\hat H_{n,m}$ to give an $H_{n,m}$ that satisfies  \eqref{lab69} for all $m\geq 1$,
write
\be \hat H_n(\bmu,\tau;\nu)=\sum_{m=0}^\infty \nu^m\hat H_{n,m}(\bmu,\tau),
\label{lab87}\ee
we have from \eqref{Gnmres},
\be\Res{\mu_n}
\hat H_n(\bmu,\tau;\nu)=\sigma(\nu,\tau)\hat H_{n-1}(\bmu',\tau;\nu).\label{lab88}\ee
So, if we define
\be {\nu^n\over\sigma(\nu,\tau)^n}\hat H_n(\bmu,\tau;\nu)
=H_n(\bmu,\tau;\nu)=\sum_{m=0}^\infty H_{n,m}(\bmu,\tau)\nu^m,\label{lab89}
\ee
the definition of $H_{n,m}$ in \eqref{explicitH} is unchanged for $1\leq m\leq 4$ (except that the constant, $k_4/24$, in the defintion of $H_{n,4}$ is determined to be $- nf_2$),
and 
\begin{align}
\Resa{\mu_n}
H_n(\bmu,\tau;\nu)&=\nu H_{n-1}(\bmu',\tau;\nu),\qquad n\geq 1,\label{Hnres}\\
\hbox{\it i.e.}\qquad
\Resa{\mu_n}
H_{n,m}(\bmu,\tau)&=H_{n-1,m-1}(\bmu',\tau).\qquad n\geq m\geq 1.\label{Hnmres}
\end{align}

From \eqref{Gnmform}
\begin{align}
\hat H_n(\bmu,\tau;\nu)&=\sum_{m=0}^\infty {\nu^m\over m!}\left[\sum_{j=1}^n(\partial_j+\zeta_j)\right]^m1\cr
&=\left[\prod_{j=1}^n{1\over\sigma(\mu_j,\tau)}\right]\sum_{m=0}^\infty {\nu^m\over m!}\left[\sum_{j=1}^n\partial_j\right]^m\prod_{j=1}^n\sigma(\mu_j,\tau)\cr
&=\prod_{j=1}^n{\sigma(\mu_j+\nu,\tau)\over\sigma(\mu_j,\tau)}\label{Gsigma}
\end{align}
and so
\be
H_n(\bmu,\tau;\nu)={\nu^n\over\sigma(\nu,\tau)^n}\prod_{j=1}^n{\sigma(\mu_j+\nu,\tau)\over\sigma(\mu_j,\tau)}.
\label{Hnform}\ee
Note that 
\be \nu^{-n}H_n(\bmu,\tau;\nu)=\prod_{j=1}^n{\sigma(\mu_j+\nu,\tau)\over\sigma(\nu,\tau)\sigma(\mu_j,\tau)}
\label{elliptic}\ee
is elliptic is a function of the $\mu_j$ and of $\nu$ provided that we impose the constraint that
$\mu_1+\mu_2+\ldots+\mu_n=0$.
From \eqref{Hnform} we see directly that
\begin{align}
\Resa{\mu_n}
H_n(\bmu,\tau;\nu)&={\nu^n\over\sigma(\nu,\tau)^n}
{\sigma(\nu,\tau)\over \sigma'(0,\tau)}\prod_{j=2}^n{\sigma(\mu_j+\nu,\tau)\over\sigma(\mu_j,\tau)}\cr
&=\nu H_{n-1}(\bmu',\tau;\nu).\label{Hnres2}
\end{align}
Properties of $H_ n(\bmu,\tau;\nu)$  are
discussed in Appendix \ref{sect:PropHn}. In particular, it is shown that
\be H_{n,n-1}(\nu_{12},\ldots,\nu_{n-1,n},\nu_{n,1},\tau)=0.\label{Hn1}\ee

The relation \eqref{Hnmres} shows that \eqref{FkH} provides the general form 
of the $n$-point connected loop amplitude, with $H_{n,m}$ defined as the 
moments of \eqref{Hnform}. It specifies $\F_n$ in terms of the invariant 
symmetric tensors $\kappa_{2,2}=\kappa$ and $\kappa_{n,0}
=\omega_n$,
\be
\F_n^{a_1a_2\ldots a_n}(\nu_1,\nu_2,\ldots, \nu_n,\tau)=
{1\over n}\sum_{\varrho\in\fS_n}\sum_{m=0}^n\kappa_{n,m}^{a_{\varrho(1)}\ldots a_{\varrho(n)}}H_{n,m}(\nu_{\varrho(1)\varrho(2)},\ldots,\nu_{\varrho(n)\varrho(1)},\tau).\label{FnkH}\ee

Since the symmetrization of $\kappa^{a_1a_2\ldots a_n}_{n,m}$ is zero for $m>0$ and $n\geq 3$, we can evaluate 
$\kappa_{n,0}$ in terms of connected parts of traces of the $J^a(\rho)$ by 
symmetrizing \eqref{FnkH}  over the group indices only, 
yielding
\be  \F^{a_1a_2\ldots a_n}(\nu_1,\nu_2,\ldots, \nu_n,\tau)_S
= (n-1)! \kappa_{n,0}^{a_1\ldots a_n}(\tau),\quad n\geq 3,\label{kn0}\ee
where we define
\be\F^{a_1a_2\ldots a_n}(\nu_1,\nu_2,\ldots, \nu_n,\tau)_S
={1\over n!}\sum_{\varrho\in\fS_n}\F_n^{a_{\varrho(1)}\ldots 
a_{\varrho(n)}}(\nu_1,\nu_2,\ldots, \nu_n,\tau).\label{lab99}\ee
The equation \eqref{kn0} can be written
\be
\omega_n^{a_1\ldots a_n}(\tau)\equiv
\kappa_{n,0}^{a_1\ldots a_n}(\tau)
=-{(2\pi i)^n\over(n-1)! \chi(\tau)}\left[\prod_{j=1}^n\rho_j\right]\tr\left(J^{a_1}(\rho_1)J^{a_2}(\rho_2)\ldots J^{a_n}(\rho_n)w^{L_0}\right)_{C,S},\label{kappan0}\ee
for $n\geq 3$; $\omega_2$ is determined by \eqref{gen2pt}.
Note that this implies that the symmetrized connected part of the trace 
\be\left[\prod_{j=1}^n\rho_j\right]\tr\left(J^{a_1}(\rho_1)J^{a_2}(\rho_2)\ldots J^{a_n}(\rho_n)w^{L_0}\right)_{C,S}\label{trCS}\ee
is independent of the $\nu_j$. (This follows directly from symmetrizing \eqref{Fres} because this shows that 
all the residues of this elliptic function vanish, implying that it is a constant on the torus.) We will relate this to the trace of zero modes of the currents in section \ref{sect:zeromodes}.
In Appendix \ref{sect:formulae} we show that the formulae given for two-, three- and four-point loops in \cite{DG} are equivalent to \eqref{FnkH} for $n\leq 4$.
\vfil\eject
\section{Zero Modes}
\label{sect:zeromodes}

\subsection{\sl Recurrence Relations and Traces of Zero Modes}
\label{sect:recurrence}

The symmetric tensor $\omega_n$ is given in terms of the symmetrized connected part of the trace of currents by \eqref{kappan0}. We seek to express this is in terms of traces of symmetrized products of zero modes, $J^a_0$. To this end consider
\be
\tr\left(J^{a_1}(\rho_1)J^{a_2}(\rho_2)\ldots J^{a_r}(\rho_r)J^{a_{r+1}}_0
\ldots J^{a_n}_0w^{L_0}\right)\prod_{j=1}^r\rho_j.\label{lab102}\ee
These functions are not elliptic as functions of the $\nu_j$, $1\leq j\leq r$, if $r<n$; to see this move 
$J^{a_1}(\rho_1)$ around the trace, through $w^{L_0}$, to calculate the effect of sending $\nu_1\rightarrow\nu_1+\tau$, and we find that it is not invariant because terms, proportional to ${f^{a_1a_j}}_e$ are generated on commuting 
$J^{a_1}(\rho_1)$ with $J^{a_j}_0$, $j>r$. However, these terms clearly disappear on symmetrizing over all the indices $(a_1,a_2,\ldots, a_n)$, so that 
\be
\tr\left(J^{a_1}(\rho_1)J^{a_2}(\rho_2)\ldots J^{a_r}(\rho_r)J^{a_{r+1}}_0
\ldots J^{a_n}_0w^{L_0}\right)_S\prod_{j=1}^r\rho_j,\label{lab103}\ee
is elliptic in $\nu_j$, $1\leq j\leq r$ and so a suitable function to consider. 

Symmetrizing the recurrence relation \cite{DG}
\begin{align}
\tr\left(J^{a_1}(\rho_1)J^{a_2}(\rho_2)\ldots J^{a_n}(\rho_n)w^{L_0}\right)&
=\rho^{-1}_1\tr\left(J^{a_1}_0J^{a_2}(\rho_2)\ldots J^{a_n}(\rho_n)w^{L_0}\right)\cr
&\hskip-54truemm+i\sum_{j=2}^n{\Delta_1(\nu_j-\nu_1,\tau)\over \rho_1}{f^{a_1a_j}}_{a_j'}\tr\left(J^{a_2}(\rho_2)\ldots J^{a_{j-1}}(\rho_{j-1})J^{a_j'}(\rho_j)J^{a_{j+1}}(\rho_{j+1})\ldots J^{a_n}(\rho_n)w^{L_0}\right)\cr
&\hskip-54truemm+k\sum_{j=2}^n{\Delta_2(\nu_j-\nu_1,\tau)\over \rho_1\rho_j}\delta^{a_1a_j}\tr\left(J^{a_2}(\rho_2)\ldots J^{a_{j-1}}(\rho_{j-1})J^{a_{j+1}}(\rho_{j+1})\ldots J^{a_n}(\rho_n)w^{L_0}\right),
\end{align} where
\be
\Delta_1(\nu,\tau)
={i\over 2\pi}{\theta_1'(\nu,\tau)\over\theta_1(\nu,\tau)}-{1\over 2},\quad
\Delta_2(\nu,\tau)
={1\over 4\pi^2}\left({\theta_1'(\nu,\tau)\over\theta_1(\nu,\tau)}\right)'=-{1\over 4\pi^2}\P(\nu,\tau)
 - {1\over 2\pi^2} \eta(\tau), \label{Delta}\ee
we obtain
\begin{align}
\tr\left(J^{a_1}(\rho_1)J^{a_2}(\rho_2)\ldots J^{a_n}(\rho_n)w^{L_0}\right)_S&
={1\over\rho_1}\tr\left(J^{a_1}_0J^{a_2}(\rho_2)\ldots J^{a_n}(\rho_n)w^{L_0}\right)_S\cr
&\hskip-50truemm+k\sum_{j=2}^n{\Delta_2(\nu_j-\nu_1,\tau)\over\rho_1\rho_j}\left[\delta^{a_1a_j}\tr\left(J^{a_2}(\rho_2)\ldots J^{a_{j-1}}(\rho_{j-1})J^{a_{j+1}}(\rho_{j+1})\ldots J^{a_n}(\rho_n)w^{L_0}\right)\right]_S.\cr
\label{JJzero}\end{align}
This generalizes to 
\begin{align}
\tr\left(J^{a_1}(\rho_1)J^{a_2}(\rho_2)\ldots J^{a_r}(\rho_r)J^{a_{r+1}}_0
\ldots J^{a_n}_0w^{L_0}\right)_S&\cr
&\hskip-72truemm={1\over\rho_1}\tr\left(J^{a_2}(\rho_2)\ldots J^{a_r}(\rho_r)J^{a_1}_0J^{a_{r+1}}_0
\ldots J^{a_n}_0w^{L_0}\right)_S
+k\sum_{j=2}^r{\Delta_2(\nu_j-\nu_1,\tau)\over\rho_1\rho_j}\cr
&\hskip-68truemm\times\left[\delta^{a_1a_j}\tr\left(J^{a_2}(\rho_2)\ldots J^{a_{j-1}}(\rho_{j-1})J^{a_{j+1}}(\rho_{j+1})\ldots J^{a_r}(\rho_r)J^{a_{r+1}}_0
\ldots J^{a_n}_0w^{L_0}\right)\right]_S.\qquad\label{Jzerorecur}
\end{align}

Applying this for $r=n=2$,
\begin{align}\tr\left(J^{a_1}(\rho_1)J^{a_2}(\rho_2)w^{L_0}\right)\rho_1\rho_2
&=\tr\left(J^{a_1}_0J^{a_2}(\rho_2)w^{L_0}\right)\rho_2
+k\Delta_2(\nu_2-\nu_1,\tau)\delta^{a_1a_2}\chi(\tau)\cr
&=\tr\left(J^{a_1}_0J^{a_2}_0w^{L_0}\right)
+k\Delta_2(\nu_2-\nu_1,\tau)\delta^{a_1a_2}\chi(\tau);
\label{rn2}\end{align} 
using \eqref{gen2pt} and \eqref{Delta}, we have
\be \omega_2^{ab} 
={4\pi^2\over\chi(\tau)}\tr\left(J^a_0J^b_0w^{L_0}\right) 
 -2 \delta^{ab} k\eta(\tau) .\label{omega2}\ee
Taking $r=n=3$,
\begin{align}
\tr\left(J^{a_1}(\rho_1)J^{a_2}(\rho_2)J^{a_3}(\rho_3)w^{L_0}\right)_S\rho_1\rho_2\rho_3
&=\tr\left(J^{a_1}_0J^{a_2}(\rho_2)J^{a_3}(\rho_3)w^{L_0}\right)_S\rho_2\rho_3\cr
&=\tr\left(J^{a_1}_0J^{a_2}_0J^{a_3}(\rho_3)w^{L_0}\right)_S\rho_3\cr
&=\tr\left(J^{a_1}_0J^{a_2}_0J^{a_3}_0w^{L_0}\right)_S
\label{rn3}\end{align}
because $\tr(J^{a_3}(\rho_3)w^{L_0})=\tr(J^{a_3}_0w^{L_0})=0$, so that
\be 
\omega_3^{abc}= {4\pi^3i\over\chi(\tau)}\tr\left(J^{a}_0J^{b}_0J^{c}_0w^{L_0}
\right)_S. \label{omega3}\ee

For $r=n=4$,
\begin{align}
\tr\left(J^{a_1}(\rho_1)J^{a_2}(\rho_2)J^{a_3}(\rho_3)J^{a_4}(\rho_4)w^{L_0}\right)_S\rho_1\rho_2
\rho_3\rho_4\hskip-75truemm&\cr
&=\tr\left(J^{a_1}_0J^{a_2}(\rho_2)J^{a_3}(\rho_3)J^{a_4}(\rho_4)w^{L_0}\right)_S\rho_2
\rho_3\rho_4\cr
&\hskip5truemm
+k\Delta_2(\nu_2-\nu_1)\left[\delta^{a_1a_2}\tr\left(J^{a_3}(\rho_3)J^{a_4}(\rho_4)w^{L_0}\right)\right]_S\rho_3\rho_4\cr
&\hskip8truemm
+k\Delta_2(\nu_3-\nu_1)\left[\delta^{a_1a_3}\tr\left(J^{a_2}(\rho_2)J^{a_4}(\rho_4)w^{L_0}\right)\right]_S\rho_2\rho_4\cr
&\hskip11truemm
+k\Delta_2(\nu_4-\nu_1)\left[\delta^{a_1a_4}\tr\left(J^{a_2}(\rho_2)J^{a_3}(\rho_3)w^{L_0}\right)\right]_S\rho_2\rho_3\cr
&=\tr\left(J^{a_1}_0J^{a_2}_0J^{a_3}(\rho_3)J^{a_4}(\rho_4)w^{L_0}\right)_S\rho_3\rho_4\cr
&\hskip5truemm
+k\left(\Delta_2(\nu_2-\nu_1)+\Delta_2(\nu_3-\nu_1)+\Delta_2(\nu_4-\nu_1)\right)
\left[\delta^{a_1a_2}\tr\left(J^{a_3}_0J^{a_4}_0w^{L_0}\right)\right]_S\cr
&\hskip8truemm
+k^2\left(\Delta_2(\nu_2-\nu_1)\Delta_2(\nu_3-\nu_4)+\Delta_2(\nu_3-\nu_1)\Delta_2(\nu_2-\nu_4)\right.\cr
&\hskip11truemm\left.+\Delta_2(\nu_4-\nu_1)\Delta_2(\nu_2-\nu_3)\right)
\left[\delta^{a_1a_2}\delta^{a_3a_4}\right]_S\chi(\tau)\cr
&\hskip14truemm
+k\Delta_2(\nu_3-\nu_2)\left[\delta^{a_2a_3}\tr\left(J^{a_1}_0J^{a_4}(\rho_4)w^{L_0}\right)\right]_S\rho_4\cr
&\hskip17truemm+k\Delta_2(\nu_4-\nu_2)\left[\delta^{a_2a_4}\tr\left(J^{a_1}_0J^{a_3}(\rho_3)w^{L_0}\right)\right]_S\rho_3\cr
&=\tr\left(J^{a_1}_0J^{a_2}_0J^{a_3}_0J^{a_4}_0w^{L_0}\right)_S\cr
&\hskip5truemm+k
\left[\delta^{a_1a_2}\tr\left(J^{a_3}_0J^{a_4}_0w^{L_0}\right)\right]_S\sum_{i<j}\Delta_2(\nu_i-\nu_j)\cr
&\hskip8truemm
+k^2\left(\Delta_2(\nu_2-\nu_1)\Delta_2(\nu_3-\nu_4)+\Delta_2(\nu_3-\nu_1)\Delta_2(\nu_2-\nu_4)\right.\cr
&\hskip11truemm\left.+\Delta_2(\nu_4-\nu_1)\Delta_2(\nu_2-\nu_3)\right)
\left[\delta^{a_1a_2}\delta^{a_3a_4}\right]_S\chi(\tau).\end{align}
Then, since
\begin{align}
\tr\left(J^{a_1}(\rho_1)J^{a_2}(\rho_2)J^{a_3}(\rho_3)J^{a_4}(\rho_4)w^{L_0}\right)_C&
=\tr\left(J^{a_1}(\rho_1)J^{a_2}(\rho_2)J^{a_3}(\rho_3)J^{a_4}(\rho_4)w^{L_0}\right)\cr
&\hskip-55truemm-\left[\tr\left(J^{a_1}(\rho_1)J^{a_2}(\rho_2)w^{L_0}\right)\tr\left(J^{a_3}(\rho_3)J^{a_4}(\rho_4)w^{L_0}\right)\right.\cr
&\hskip-45truemm\left.+\tr\left(J^{a_1}(\rho_1)J^{a_3}(\rho_3)w^{L_0}\right)\tr\left(J^{a_2}(\rho_2)J^{a_4}(\rho_4)w^{L_0}\right)\right.\cr
&\hskip-35truemm\left.+\tr\left(J^{a_1}(\rho_1)J^{a_4}(\rho_4)w^{L_0}\right)\tr\left(J^{a_2}(\rho_2)J^{a_3}(\rho_3)w^{L_0}\right)\right]
/\chi(\tau),
\end{align}
\begin{align}
\tr\left(J^{a_1}(\rho_1)J^{a_2}(\rho_2)J^{a_3}(\rho_3)J^{a_4}(\rho_4)w^{L_0}\right)_{CS}\prod_{j=1}^4\rho_j&\cr
&\hskip-65truemm
=\tr\left(J^{a_1}_0J^{a_2}_0J^{a_3}_0J^{a_4}_0w^{L_0}\right)_S
-3\left[\tr\left(J^{a_1}_0J^{a_2}_0w^{L_0}\right)\tr
\left(J^{a_3}_0J^{a_4}_0w^{L_0}\right)\right]_S/\chi(\tau).\qquad\label{omega4}
\end{align}
From \eqref{kappan0}, this shows that $\omega_4$ is given as a ``connected part'' of a trace of zero modes. In the next section we define such connected parts and show that $\omega_n$ is given in terms of them for all $n\geq 2$.

\subsection{\sl Connected Parts of Zero Mode Amplitudes}
\label{connectzero}

Because of the locality of the currents, $\A^A$, defined as in \eqref{AA}, and so $\A^A_C$, defined
inductively by \eqref{inductproc}, is symmetric under simultaneous permutations of the indices $a_j$ and the variables $\nu_j$. We define the symmetrization $\A^A_S$ of $\A^A$ by symmetrizing on the $a_j$ alone:
\be\A^{a_{i_1}a_{i_2}\ldots a_{i_n}}_S(\nu_{i_1},\nu_{i_2},\ldots,\nu_{i_n},\tau)
={1\over n!}\sum_{\varrho\in\fS_n}\A^{a_{i_{\varrho(1)}}a_{i_{\varrho(2)}}\ldots a_{i_{\varrho(n)}}}(\nu_{i_1},\nu_{i_2},\ldots,\nu_{i_n},\tau);\label{Symmetrization}
\ee
equivalently we could symmetrize on the variables $\nu_j$ alone. We define $\A^A_{CS}$, the symmetrization of $\A^A_C$, similarly.

We consider the trace of zero modes of the currents, 
\be\Z^A\equiv \Z^{a_{i_1}a_{i_2}\ldots a_{i_n}}(\tau)
=\tr\left(J^{a_{i_1}}_0J^{a_{i_2}}_0\ldots J^{a_{i_n}}_0w^{L_0}\right)
(2\pi i)^n,\label{lab124}\ee
and, more particularly, its symmetrization, $\Z^A_S$, defined as in \eqref{Symmetrization}.
We can define a "connected part", $\Z_{CS}$,  inductively for $\Z^A_S$, following \eqref{inductproc},
\be\Z^A_{CS}=\Z^A_S-\sum_{P\in\fP_A'}\chi(\tau)^{1-|P|}\prod_{A_j\in P}\Z^{A_j}_{CS},\label{ZCSZS}\ee
(where again $\fP_A'$ denotes the same collection of divisions of $A$ into disjoint subsets but omitting the division of $A$ into the single set consisting of itself) together with the vanishing of the one point function $\Z^{\{i\}}_{CS}=0$, and with the two-point function given by
\be\Z^{\{i,j\}}_{CS}=\Z^{\{i,j\}}_S=\tr\left(J^{a_{i}}_0J^{a_{j}}_0w^{L_0}\right)
(2\pi i)^2.\label{lab126}\ee

For $A=\{i_1,i_2,\ldots,i_{2m}\}$, define
\be\P^A={k^m(2\pi i)^2\over 2^mm!}\sum_{\varrho\in\fS_{2m}}\prod_{j=1}^m\delta^{a_{i_{\varrho(2j-1)}}a_{i_{\varrho(2j)}}}
\Delta_2(\nu_{i_{\varrho(2j-1)}}-\nu_{i_{\varrho(2j)}})\label{lab128}\ee
and define $\P^A=0$ if $A$ has an odd number of elements. Then
\be\A^{\{i,j\}}_{CS}=\Z^{\{i,j\}}_{CS} +\P^{\{i,j\}}\chi,\label{ACSZCS}\ee
and the recurrence relation \eqref{Jzerorecur} leads to
\be\A^A_S=\Z^A_S+\sum_{B\in\fR_A}[\P^B\Z^{A\sim B}_S]_S\label{recAAS}\ee
where $\fR_A$ denotes the subsets of $A$, excluding the empty set but 
including $A$ itself. 

Now, symmetrizing \eqref{inductproc},
\be\A^A_S=\A^A_{CS}+\sum_{P\in\fP_A'}\chi^{1-|P|}\left[\prod_{A_j\in P}\A^{A_j}_{CS}\right]_S.\label{lab131}\ee
If we assume, as the inductive hypothesis, that $\A^B_{CS}=\Z^B_{CS}$, 
for $2<|B|<|A|$, and that \eqref{ACSZCS} holds when $|B|=2$, we have, 
on substituting for $\A^{A_j}_{CS}$ and symmetrizing, that
\begin{align}
\A^A_S&=\A^A_{CS}+\sum_{P\in\fP_A'}\chi^{1-|P|}\left[\prod_{A_j\in P}\Z^{A_j}_{CS}\right]_S
+\sum_{B\in\fR_A}\left[\P^B\sum_{R\in\fP_{A\sim B}}\chi^{1-|R|}\prod_{D_j\in R}\Z^{D_j}_{CS}\right]_S\cr
&=\A^A_{CS}+\sum_{P\in\fP_A'}\chi^{1-|P|}\left[\prod_{A_j\in P}\Z^{A_j}_{CS}\right]_S
+\sum_{B\in\fR_A}\left[\P^B\Z^{A\sim B}\right]_S\end{align}
by \eqref{ZCSZS}.
Then using \eqref{recAAS}, 
\be\Z^A_S=\A^A_{CS}+\sum_{P\in\fP_A'}\chi^{1-|P|}\left[\prod_{A_j\in P}\Z^{A_j}_{CS}\right]_S,\label{lab132}\ee
so that $\A^A_{CS}$ satisfies the recurrence relation \eqref{ZCSZS} for $\Z^A_{CS}$ and we can conclude inductively that $\A^A_{CS}=\Z^A_{CS}$, for $|A|>2$.

It follows from \eqref{kappan0},
\be
\omega_n(\tau)  =
\kappa_{n,0}^{a_1\ldots a_n}(\tau)
=-{(2\pi i)^n\over(n-1)!\chi(\tau)}\tr\left(J^{a_1}_0J^{a_2}_0\ldots J^{a_n}_0w^{L_0}\right)_{C,S}, \quad n\geq 3,\label{lab133}\ee
with $\omega_2$ given by \eqref{omega2}.

\subsection{\sl Traces of Zero Modes and Characters}
\label{sect:zeromodesch}

In this section we will relate the symmetrized traces of the zero modes 
\be \Z^{a_{i_1}a_{i_2}\ldots a_{i_n}}_S(\tau)
=\tr\left(J^{a_{i_1}}_0J^{a_{i_2}}_0\ldots J^{a_{i_n}}_0w^{L_0}\right)_S
(2\pi i)^n,\label{lab157a}\ee
to the character of the representation of the affine algebra, $\hat\g$, defined by \eqref{alg}, in the space of states,
\be\chi(\theta,\tau)=\tr\left(e^{i H\cdot\theta} w^{L_0}\right).\label{chidef1}\ee
Here $H$ denotes the generators of a Cartan subalgebra, $\h$, of the finite-dimensional algebra $\g$ formed by the zero modes,
\be[J^a_0,J^a_0]={f^{ab}}_cJ^c_0.\label{lab159}\ee
For convenience of exposition, we shall take $\g$ to be simple in what follows. 

For fixed $\tau$, $\tr\left(J^{a_{i_1}}_0J^{a_{i_2}}_0\ldots J^{a_{i_n}}_0w^{L_0}\right)_S$ 
is an invariant symmetric  tensor for $\g$, The space of symmetric tensors, $\Sc(\g)$ is isomorphic 
(as a vector space) to $\U(\g)$, the universal enveloping algebra of $\g$,
\be\omega^{a_1a_2\ldots a_n}\rightarrow\omega^{a_1a_2\ldots a_n}J^{a_{i_1}}_0J^{a_{i_2}}_0\ldots J^{a_{i_n}}_0. \label{lab160}\ee
The invariant tensors  $\Sc(\g)^\g\subset\Sc(\g)$ correspond to the center $Z(\U(\g))$ of $\U(\g)$, {\it i.e.} the elements of $\U(\g)$ that commute with $\g$. This is a ring generated by rank $\g$ elements 
({\it e.g.} \cite{VSV}, page 337), the basic Casimir operators, or primitive invariant tensors. These can be taken to be orthogonal,
\be\omega^{a_1a_2\ldots a_n}\omega_{a_1a_2\ldots a_m}'=0,\label{lab161}\ee
where $\omega, \omega'$ are primitive invariant symmetric tensors of orders $n, m$, $m<n$.

We can use a Cartan-Weyl basis for $\g$, using $\Phi$ to denote the set of roots of $\g$,
\begin{align}[H^i,H^j]&=\hbox to23truemm{$0,$\hfill}\qquad 1\leq i,j\leq \hbox{rank}\g;\cr
[H^i,E^\alpha]&=\hbox to23truemm{$\alpha^iE^\alpha,$\hfill}\qquad \alpha\in\Phi,\quad 1\leq i\leq\hbox{rank }\g;\cr
[E^\alpha,E^\beta]&=\hbox to23truemm{$\epsilon(\alpha,\beta)E^{\alpha+\beta},$\hfill}\qquad\alpha,\beta,\alpha+\beta\in\Phi\cr
&=\hbox to23truemm{$\displaystyle {2\over \alpha^2}\alpha\cdot H,$\hfill}\qquad \beta=-\alpha\in\Phi\cr
&=\hbox to23truemm{$0,$\hfill}\qquad\hbox{otherwise.\hfil}\label{CWbasis}
\end{align}
(We omit the suffix $0$ on $H, E^\alpha$.) With this choice of basis, the quadratic Casimir operator
\begin{align}
J^aJ^a&= H^2+\sum_{\alpha>0}{\alpha^2\over 2}\left(E_{-\alpha}E_\alpha+E_\alpha E_{-\alpha}\right)\cr
&= H^2+2\delta\cdot H+\sum_{\alpha>0}\alpha^2E_{-\alpha}E_\alpha, 
\qquad \delta={1\over 2}\sum_{\alpha>0}\alpha,\cr
&= (H+\delta)^2- \delta^2+\sum_{\alpha>0}\alpha^2E_{-\alpha}E_\alpha.\label{QCasimir}\end{align}
The value of $J^2$ in a representation with highest weight $\lambda$ can be obtained by evaluating this on the highest weight state $|\lambda\rangle$, which has $E^\alpha|\lambda\rangle=0$ for $\alpha>0$. Thus the value of $J^2$ in this representation is 
\be\lambda\cdot(\lambda+2\delta)=\left(\lambda+\delta\right)^2-\delta^2.\label{lab164}\ee

If $\xi^{a_1a_2\ldots a_n}$ is any invariant tensor for $\g$, 
\be C_\xi=\xi^{a_1a_2\ldots a_n}J^{a_1}_0J^{a_2}_0\ldots J^{a_n}_0\in Z(\U(g)),\label{lab165}\ee
and we can evaluate its value in the representation with highest weight $|\lambda\rangle$ by expressing it in the Cartan-Weyl basis and moving the $E_\alpha, \alpha >0,$ to the right. Because $[H^i,C_\beta]=0$, we can write
\be C_\xi=\phi_\xi(H)+\sum_{\alpha>0}  F_{\xi,\alpha}E_\alpha, \qquad\hbox{for suitable } 
F_{\xi,\alpha}\in\U(\g),\label{lab166}\ee
where $\phi_\xi(H)$ is a polynomial of degree $n$ in the $H^i$. Then $C_\xi|\lambda\rangle=
\phi_\xi(\lambda)|\lambda\rangle$, so that $C_\xi$ takes the value $\phi_\xi(\lambda)$ in the representation with highest weight $\lambda$. Given two such invariant tensors $\xi_1, \xi_2$,
\begin{align}
C_{\xi_1}C_{\xi_2}&=\left(\phi_{\xi_1}(H)+\sum_{\alpha>0}  F_{\xi_1,\alpha}E_\alpha\right)\left(\phi_{\xi_2}(H)+\sum_{\beta>0}  F_{\xi_2,\beta}E_\beta\right)\cr
&=\phi_{\xi_1}(H)\phi_{\xi_2}(H)+\sum_{\alpha>0}  F_{\xi_1,\alpha}\phi_{\xi_2}(H-\alpha)E_\alpha
+\sum_{\beta>0}  \phi_{\xi_1}(H)F_{\xi_2,\beta}E_\beta\cr
&\hskip50truemm+\sum_{\alpha>0}  F_{\xi_1,\alpha}\phi_{\xi_2}(H)E_\alpha\sum_{\beta>0}  \phi_{\xi_1}(H)F_{\xi_2,\beta}E_\beta
\end{align}
so that 
\be\phi_{\xi_1\xi_2}(H)=\phi_{\xi_1}(H)\phi_{\xi_2}(H).\label{lab167}\ee

If $\phi_{\xi_1}=\phi_{\xi_2}$, then $C_{\xi_1}=C_{\xi_2}$ acting in each highest weight representation of $\g$. It follows from this that $C_{\xi_1}=C_{\xi_2}$ as elements of $\U(\g)$ (see, e.g., \cite{AWK}, page 251). Thus $C_\xi\mapsto \phi_\xi$ defines a map $Z(\U(\g))\rightarrow \Sc(\h)$, which is an algebra homomorphism and is one-to-one. 

The elements $\phi_\xi\in\Sc(\h)$ obtained in this way have an invariance under the Weyl group, $W$, of $\g$ as we shall now show (see \cite{JEH}, page 130, or \cite{AWK}, page 246). Consider the action of $\phi_\xi$ in the infinite-dimensional representation, $\tilde V_\lambda$, with highest weight $\lambda$, where $\lambda\in\Lambda_\g$, the weight lattice of $\g$, with $\alpha\cdot\lambda\geq 0$ for $\alpha>0$, whose states are generated by the action of $E_\alpha, \alpha >0,$ on a state $|\lambda\rangle$. The finite-dimensional representation, $V_\lambda$, is the quotient of $\tilde V_\lambda$ by its largest invariant subspace. Taking a basis of simple roots, $\alpha_1,\alpha_2,\ldots, \alpha_r$, $r=\hbox{rank }\g$, $m_i=2\alpha_i\cdot\lambda/\alpha_i^2\in\Zop$ and  $\alpha_i\cdot\lambda\geq 0$, consider $E_{-\alpha_i}^{m_i+1}|\lambda\rangle$. Now
$$\alpha_i\cdot HE_{-\alpha_i}^{s}|\lambda\rangle=\half \alpha_i^2(m_i-2s)E_{-\alpha_i}^{s}|\lambda\rangle$$
so
$$E_{\alpha_i}E_{-\alpha_i}^{m_i+1}|\lambda\rangle=\sum_{s=0}^{m_i}(m_i-2s)E_{-\alpha_i}^{m_i}|\lambda\rangle=0.$$
and $E_{\alpha_j}E_{-\alpha_i}^{m_i+1}|\lambda\rangle=0$ for $i\ne j$ because $[E_{\alpha_i},E_{\alpha_j}]=0$, $i\ne j$. It follows that $E_\alpha E_{-\alpha_i}^{m_i+1}|\lambda\rangle=0$ for $\alpha>0$ and so $ E_{-\alpha_i}^{m_i+1}|\lambda\rangle=0$ generate an invariant subspace of $\tilde V_\lambda$ (which is divided out in the construction of $V_\lambda$). Then 
\be C_\xi E_{-\alpha_i}^{m_i+1}|\lambda\rangle=\phi_\xi(H)E_{-\alpha_i}^{m_i+1}|\lambda\rangle
=\phi_\xi(\lambda-m_i\alpha_i-\alpha_i)E_{-\alpha_i}^{m_i+1}|\lambda\rangle.\label{lab168a}\ee
But, on the other hand 
\be C_\xi E_{-\alpha_i}^{m_i+1}|\lambda\rangle=E_{-\alpha_i}^{m_i+1}|C_\xi \lambda\rangle=
\phi_\xi(\lambda) E_{-\alpha_i}^{m_i+1}|\lambda\rangle.\label{lab168b}\ee
Thus, for each simple root, $\alpha_i$,
\be \phi_\xi(\lambda)=\phi_\xi(\lambda-m_i\alpha_i-\alpha_i).\label{phixilambda}\ee
If $\sigma_i$ denotes the element of the $W_\g$ corresponding to reflection in the hyperplane orthogonal to $\alpha_i$, 
$$\sigma_i(\lambda)=\lambda-m_i\alpha_i, \qquad\hbox{and}\qquad 
\sigma_i(\delta)=\delta-\alpha_i,$$
because $2\delta\cdot\alpha_i/\alpha_i^2=1$ for each simple root $\alpha_i$. Thus \eqref{phixilambda} can be rewritten
\be\phi_\xi(\lambda)=\phi_\xi(\sigma_i(\lambda+\delta)-\delta),\label{lab170}\ee
and, if we define
\be\tilde\phi_\xi(\lambda)=\phi_\xi(\lambda-\delta)=\phi_\xi(\sigma_i(\lambda)-\delta)
=\tilde\phi(\sigma_i(\lambda)).\label{lab171}\ee
Because the reflections in the simple roots, $\sigma_i$, generate the Weyl group $W_\g$, $\tilde\phi_\xi(\lambda)=\phi_\xi(\lambda-\delta)$ defines a function invariant under the whole Weyl group. Thus
$C_\xi\mapsto \tilde\phi_\xi$  defines a homomorphism of $Z(\U(\g))\rightarrow \Sc(\h)^W,$ the polynomials in $H$ invariant under the Weyl group. In fact, this map is an isomorphism, called the Harish-Chandra isomorphism. That $C_\xi\mapsto \tilde\phi_\xi$ is onto follows from the fact that
$\Sc(\h)^W$ is spanned by $\phi_\xi$ for $\xi^{a_1a_2\ldots a_n}=\tr(t^{a_1}t^{a_2}\ldots t^{a_n})$, where the $t^a$ are the representations of $J^a_0$ in the finite-dimensional representation $V_\lambda,\lambda\in\Lambda_\g$  (see, e.g., \cite{AWK}, page 253).

Now, writing
\be\tr\left(e^{i H\cdot\theta} w^{L_0}\right)=\chi(\theta,\tau)=\sum_{\lambda\in\Lambda_\g^+}b_\lambda(w)\chi^\lambda(\theta),\label{lab172}\ee
where $\Lambda^+_\g=\{\lambda\in\Lambda_\g:\alpha\cdot\lambda\geq 0\hbox{ for }\alpha>0\}$,
\be\tr\left(C_\xi w^{L_0}\right)=\sum_{\lambda\in\Lambda_\g^+}b_\lambda(w)\tilde\phi(\lambda+\delta)\dim V_\lambda,\label{Casxi}\ee
and the character for the finite-dimensional representation $V_\lambda$ of $\g$, $\chi^\lambda(\theta) $ is given by the Weyl character formula,
\be\chi^\lambda(\theta)={1\over \Delta_\g(\theta)}\sum_{\sigma\in W_\g} \epsilon(\sigma)
e^{i\sigma(\delta+\lambda)\cdot\theta},\label{WChar}\ee
with $\epsilon(\sigma)=\pm 1$ being the determinant of $\sigma$, and  the Weyl denominator being given by 
\be\Delta_\g(\theta)=\prod_{\alpha>0}\left(e^{{i\over2}\alpha\cdot\theta}-e^{-{i\over2}\alpha\cdot\theta}\right),\label{WDenom}\ee
where the product is over the positive roots of $\g$ (See \cite{JEH}, page 139.) The dimension $\dim V_\lambda=\chi^\lambda(0)$, but to evaluate this from \eqref{WChar}, we need to take a limit on the right hand side. In fact
$\Delta_\g(\theta)={\cal O}(\theta^{n^+})$, as $\theta\rightarrow 0$, where $n^+$ is the number of positive roots of $\g$. Now
\begin{align}
\prod_{\alpha>0}\alpha\cdot\partial_\theta\sum_{\sigma\in W_\g} \epsilon(\sigma)
e^{i\sigma(\delta+\lambda)\cdot\theta}&=\prod_{\alpha>0}\sum_{\sigma\in W_\g} i\epsilon(\sigma)
e^{i\sigma(\delta+\lambda)\cdot\theta}\sigma(\delta+\lambda)\cdot\alpha\cr
&=\prod_{\alpha>0}\sum_{\sigma\in W_\g} i\epsilon(\sigma)
(\delta+\lambda)\cdot\sigma^{-1}(\alpha),\qquad\hbox{when }\theta=0.\label{alphath}
\end{align}
As $\alpha$ runs over the positive roots of $\g$, $\sigma(\alpha)$ will range over a set obtained from the positive roots by reversing some of their signs. The product of these sign changes equals $\epsilon(\sigma)=\epsilon(\sigma^{-1})$. Hence the sign changes cancel the effect of $\epsilon(\sigma)$ in \eqref{alphath} and we have
\be\left.\prod_{\alpha>0}\alpha\cdot\partial_\theta\sum_{\sigma\in W_\g} \epsilon(\sigma)
e^{i\sigma(\delta+\lambda)\cdot\theta}\right|_{\theta=0}=i^{n^+}|W_\g|\prod_{\alpha>0}(\delta+\lambda)\cdot\alpha.\label{lab176}\ee
Since $\chi^0(\theta)=1$,
\be\Delta_\g(\theta)=\sum_{\sigma\in W_\g} \epsilon(\sigma)
e^{i\sigma(\delta)\cdot\theta},\label{lab177}\ee
and, hence,
\be\left.\prod_{\alpha>0}\alpha\cdot\partial_\theta\Delta_\g(\theta)\right|_{\theta=0}=i^{n^+}|W_\g|\prod_{\alpha>0}\delta\cdot\alpha,\label{lab178}\ee
and
\be\dim V_\lambda=\chi^\lambda(0)=\prod_{\alpha>0}{(\delta+\lambda)\cdot\alpha\over\delta\cdot\alpha}.\label{lab179}\ee

Applying
\be(\beta\cdot\partial_\theta)^n\prod_{\alpha>0}\alpha\cdot\partial_\theta \label{lab180}\ee 
to the equation
\be\chi^\lambda(\theta) \Delta_\g(\theta)=\sum_{\sigma\in W_\g} \epsilon(\sigma)
e^{i\sigma(\delta+\lambda)\cdot\theta},\label{lab181}\ee
we obtain
\begin{align}
\left.(\beta\cdot\partial_\theta)^n\left(\prod_{\alpha>0}\alpha\cdot\partial_\theta\right)
\chi^\lambda(\theta) \Delta_\g(\theta)\right|_{\theta=0}&=i^{n^++n}\prod_{\alpha>0}(\delta+\lambda)\cdot\alpha\sum_{\sigma\in W_\g} 
\left(\sigma(\delta+\lambda)\cdot\beta\right)^n\cr
&=i^{n^++n}\prod_{\alpha>0}\delta\cdot\alpha\sum_{\sigma\in W_\g} 
\left(\sigma(\delta+\lambda)\cdot\beta\right)^n\dim V_\lambda\label{lab182}
\end{align}
and so
\be\left.(\beta\cdot p_\theta)^n\left(\prod_{\alpha>0}{\alpha\cdot p_\theta\over \alpha\cdot\delta}\right)
\chi(\theta,\tau) \Delta_\g(\theta)\right|_{\theta=0}=\sum_\lambda b_\lambda(w)\sum_{\sigma\in W_\g} 
\left(\sigma(\delta+\lambda)\cdot\beta\right)^n\dim V_\lambda.\label{lab183}\ee
where $p_\theta=-i\partial_\theta$.

If 
\be\tilde\phi(\lambda)=\sum_{\sigma'\in W_\g}\left(\beta\cdot\sigma'(\lambda)\right)^n=\sum_{\sigma'\in W_\g}\left(\sigma'(\beta)\cdot\lambda\right)^n,\label{phitilde}\ee
\begin{align}
\left.\tilde\phi( p_\theta)\left(\prod_{\alpha>0}{\alpha\cdot p_\theta\over \alpha\cdot\delta}\right)
\chi(\theta,\tau) \Delta_\g(\theta)\right|_{\theta=0}&=\sum_\lambda b_\lambda(w)\sum_{\sigma,\sigma'\in W_\g} 
\left(\sigma(\delta+\lambda)\cdot\sigma'(\beta)\right)^n\dim V_\lambda\cr
&=|W_\g|\sum_\lambda b_\lambda(w) 
\tilde\phi(\delta+\lambda)\dim V_\lambda.\label{phitilde2}
\end{align}
The functions \eqref{phitilde} span the polynomial functions $\tilde\phi(\lambda)$ invariant under the Weyl group and so \eqref{phitilde2} holds for any such function. From \eqref{Casxi}, it follows that
\be\tr\left(C_\xi w^{L_0}\right)=\left.{1\over |W_\g|}\tilde\phi_\xi(p_\theta)
\left(\prod_{\alpha>0}{\alpha\cdot p_\theta\over \alpha\cdot\delta}\right)
\chi(\theta,\tau) \Delta_\g(\theta)\right|_{\theta=0}.\label{lab186}\ee

The symmetrized products of the primitive symmetric invariant tensors form a basis for all symmetric invariant tensors. Suppose $\omega^{a_1a_2\ldots a_n}_j, 1\leq j\leq N$ forms an orthonormal basis for the symmetric invariant tensors of order $n$, so that 
\be\omega^{a_1a_2\ldots a_n}_j\omega^{a_1a_2\ldots a_n}_k=\delta_{jk}.\label{lab187}\ee
Then we can write
\be\tr\left(J^{a_{i_1}}_0J^{a_{i_2}}_0\ldots J^{a_{i_n}}_0w^{L_0}\right)_S
=\sum_{j=1}^Nf_j(w)\omega^{a_1a_2\ldots a_n}_j.\label{lab188}\ee
where
\be f_j(w)=\tr\left(C_{\omega_j}w^{L_0}\right)\label{lab189}\ee
and
\begin{align}
\tr\left(J^{a_{i_1}}_0J^{a_{i_2}}_0\ldots J^{a_{i_n}}_0w^{L_0}\right)_S\hskip-25truemm&\cr
&={1\over |W_\g|}\sum_{j=1}^N\omega^{a_1a_2\ldots a_n}_j
\left.\tilde\phi_{\omega_j}(p_\theta)
\left(\prod_{\alpha>0}{\alpha\cdot p_\theta\over \alpha\cdot\delta}\right)
\chi(\theta,\tau) \Delta_\g(\theta)\right|_{\theta=0}\cr
&={1\over |W_\g|}\sum_{j=1}^N\omega^{a_1a_2\ldots a_n}_j\sum_{\sigma\in W_\g}\epsilon(\sigma)
\left.\tilde\phi_{\omega_j}(p_\theta+\sigma(\delta))
\left(\prod_{\alpha>0}{\alpha\cdot (p_\theta+\sigma(\delta))\over \alpha\cdot\delta}\right)
\chi(\theta,\tau) \right|_{\theta=0}\cr
&={1\over |W_\g|}\sum_{j=1}^N\omega^{a_1a_2\ldots a_n}_j\sum_{\sigma\in W_\g}
\left.\phi_{\omega_j}(\sigma(p_\theta))
\left(\prod_{\alpha>0}{\alpha\cdot (\delta+\sigma(p_\theta))\over \alpha\cdot\delta}\right)
\chi(\theta,\tau) \right|_{\theta=0}\label{lab190b}
\end{align}
\vfil\eject
\section{Summary and Conclusions}
\label{conclusions}

In this paper, we have constructed a general formula for the loop amplitude 
\be\tr\left(J^{a_1}(\rho_1)J^{a_2}(\rho_2)\ldots J^{a_n}(\rho_n)w^{L_0}\right),\label{Aloop3}\ee
where the currents $J^a(\rho)$ satisfy the operator product expansion 
\be
J^a(z_1)J^b(z_2)\sim{\kappa^{ab}\over (z_1-z_2)^2}+ {{f^{ab}}_cJ^c(z_2)\over z_1-z_2},
\label{ope3}
\ee 
which is equivalent to the affine algebra $\hat\g$, defined by \eqref{alg}. This formula extends the Frenkel-Zhu construction for tree amplitudes \cite{FZ}, described in section \ref{sect:FZ}, and generalizes the results obtained when the currents are given as bilinear  expressions in fermionic fields, which are reviewed in \ref{fermionloop}.

The general formula is described  graphically  by summing over all graphs with $n$ vertices where the  vertices carry the labels $a_1,a_2,\ldots, a_n$ and each vertex is connected by directed lines to other vertices, one of the lines at each vertex pointing towards it and one away from it.  Each graph consists of a number, $r$, of directed ``loops'' or cycles, $\xi=(i_1,i_2\ldots i_\ell)$ with which we associate an expression
$f_\xi$. The expression associated with the whole graph consists of a factor of $1/2\pi i\rho_j$ for each current $J^{a_j}(\rho_j)$ and $-f_{\xi_i}, 1\leq i\leq r$, for each cycle,
\be \left[\prod_{j=1}^n{1\over 2\pi i\rho_j}\right]\sum_{\hbox{\tiny diagrams}}
(-1)^r\prod_{i=1}^rf_{\xi_i}\label{general}\ee
For $\xi=(i_1,i_2\ldots i_\ell)$,
\be
f_\xi=\sum_{m=0}^\ell\kappa_{\ell,m}^{a_{i_1}a_{i_2}\dots a_{i_\ell}}H_{\ell,m}
(\nu_{{i_1}{i_2}},\dots , \nu_{i_{\ell-1} i_\ell}, \nu_{i_\ell i_1},\tau).
\ee
The functions $H_{\ell,m}$ are defined in terms of the Weierstrass $\sigma$ function by 
\be
{\nu^\ell\over\sigma(\nu,\tau)^\ell}\prod_{j=1}^\ell{\sigma(\mu_j+\nu,\tau)\over\sigma(\mu_j,\tau)}
=\sum_{m=0}^\infty H_{\ell,m}(\mu_1,\mu_2,\ldots,\mu_\ell,\tau)\nu^m,
\label{Hnform2}\ee
and the invariant tensors $\kappa_{\ell,m}$ are defined inductively by the equations 
\be\kappa_{\ell,m}^{a_1a_2\ldots a_\ell}(\tau)-\kappa_{\ell,m}^{a_2a_1\ldots a_\ell}(\tau)
={f^{a_1a_2}}_b\kappa_{\ell-1,m-1}^{ba_3\ldots a_\ell}(\tau),\label{kappatau12}\ee
\be\kappa_{\ell,m}^{a_1a_2\ldots a_\ell}(\tau)=\kappa_{\ell,m}^{a_2\ldots a_{\ell}a_{1}}(\tau),\label{kappatau22}\ee
together with the requirement that $\kappa_{\ell,m}$ be orthogonal to all symmetric tensors for $m>0$ and $\ell>2$, and the initial condition that $\kappa_{2,2}=\kappa, \kappa_{2,1}=0$ and $\kappa_{\ell,0}(\tau)=\omega_\ell(\tau)$, where the symmetric invariant tensor,
\be \omega_\ell^{a_1a_2\ldots a_\ell}(\tau)
 =-{(2\pi i)^\ell\over  (\ell-1)!\chi(\tau)} 
\tr\left(J^{a_1}_0J^{a_2}_0\ldots J^{a_\ell}_0w^{L_0}\right)_{C,S}, \quad \ell\geq 3;\label{omegaJ}\ee
\be 
\omega_2^{ab}  ={4\pi^2\over\chi(\tau)}\tr\left(J^a_0J^b_0w^{L_0}\right)
 -2 \delta^{ab} k \eta(\tau) .\label{omega5}\ee
 A proof that $\kappa_{\ell,m}$ is exists and is defined uniquely by \eqref{kappatau12} and \eqref{kappatau22} is given in section \ref{sect:Tensors} and an algorithmic method for constructing them inductively using Young tableaux is given in Appendix \ref{sect:constrkappa}.
 
 The results described so far in this section apply to the affine algebra, $\hat\g$,  associated with any finite-dimensional Lie algebra, $\g$. In section \ref{sect:zeromodesch} we gave a method for calculating the traces of zero modes,  necessary to determine the symmetric tensors $\omega$, in terms of the character 
 \be\chi(\theta,\tau)=\tr\left(e^{i H\cdot\theta} w^{L_0}\right)\label{chidef2}\ee
 of the representation provided by the space of states of the theory. The method would apply to any compact $\g$ but we took it to be simple for ease of exposition. 
 
 The ``connected'' symmetrized trace 
 \eqref{omegaJ} is defined in section \ref{connectzero} in terms of the ``full'' symmetrized traces, which themselves can be expanded in terms of an orthonormal basis of symmetric invariant tensors of order $\ell$,
 \be\tr\left(J^{a_{i_1}}_0J^{a_{i_2}}_0\ldots J^{a_{i_\ell}}_0w^{L_0}\right)_S
=\sum_{j=1}^Nf_j(w)\omega^{a_1a_2\ldots a_\ell}_j,\label{omegaexp}\ee
where $N$ is the number of independent symmetric invariant tensors of order $\ell$,
$f_j(w)=\tr\left(C_{\omega_j}w^{L_0}\right)$ and the Casimir operator
 $C_{\omega_j}=\omega_j^{a_1a_2\ldots a_\ell}J^{a_1}_0J^{a_2}_0\ldots J^{a_\ell}_0$. In section \ref{sect:zeromodesch}, we reviewed how a normal ordering of the $J^a_0$ in $C_{\omega_j}$, by writing 
 $C_{\omega_j}=\phi_{\omega_j}(H) + N_j$, where $H$ denotes the elements of a Cartan subalgebra and $N_j$ annihilates highest weight states, so that $C_{\omega_j}\mapsto\phi_{\omega_j}(H)$
 defines the Harish-Chandra isomorphism of the center of the enveloping algebra of $\g$ (that is the ring of Casimir operators) onto the polynomials in $H$ invariant under the action of the Weyl group, $W_\g$ of $\g$. This leads to the expression
\be {1\over |W_\g|}\sum_{j=1}^N\omega^{a_1a_2\ldots a_n}_j\sum_{\sigma\in W_\g}
\left.\phi_{\omega_j}(\sigma(p_\theta))
\left(\prod_{\alpha>0}{\alpha\cdot (\delta+\sigma(p_\theta))\over \alpha\cdot\delta}\right)
\chi(\theta,\tau) \right|_{\theta=0}\ee
for the symmetrized trace \eqref{omegaexp}, where $p_\theta
=-i\partial_\theta$, $\alpha$ denotes a root of $\g$ and $\delta$ denotes 
half the sum of positive roots. With this we have assembled all the elements 
of an explicit expression for the loop amplitude \eqref{Aloop3}.
In Appendix C, this is compared with expressions given previously \cite{DG} for $n=2,3,4$.

\section*{Acknowledgements}

We are grateful to Matthias Gaberdiel for helpful correspondence.
LD thanks the Institute for Advanced Study at Princeton for its hospitality,
and was partially supported by the U.S. Department of Energy,
Grant No. DE-FG01-06ER06-01, Task A.
\vfill\eject

\appendix
\section{Explicit Construction of the Tensor $\kappa_n$ in terms of $\kappa_{n-1}$}
\label{sect:constrkappa}

As a preparation for giving an explicit construction of the tensor $\kappa_n$ in terms of $\kappa_{n-1}$, we review some salient features of the representation theory of $\fS_n$ (see {\it e.g.} \cite{FH}, page 44).

The number of inequivalent irreducible representations of $\fS_n$, the group of permutations of $n$ objects, is $p(n)$, the number of partitions of $n$. Each partition, $p=(p^1,\ldots, p^m)$,  with $p^i\geq p^j$, if $i\leq j$ and $\sum_{i=1}^mp^i=n$, determines a Young diagram, consisting of $n$ boxes arranged into $m$ rows and $p^1$ columns, with $p^i$ boxes in the $i$-th row and the number of boxes in the $j$-th column equal to the number of $p^k\geq j$. We can identify the partition $p$ with the corresponding Young diagram. The Young diagrams label the inequivalent irreducible representations. 

Given a Young diagram $p$, a Young tableau, $\lambda$, is defined by an assignment of 
the integers $1,\ldots, n$ to the $n$ boxes of $p$. This gives $n!$ Young tableaux associated with a given Young diagram.
A standard Young tableau is one for which the numbers assigned to the boxes decrease along each row (from left to right) and down each column. The number of standard Young tableau associated with the Young diagram $p$,
\be d_p = {n!\over \ell_1!\ldots\ell_m!}\prod_{i<j}(\ell_i-\ell_j),\qquad\hbox{where } \ell_j=p^j+m-j,
\label{dimlambda}\ee
and this is also the dimension of the irreducible representation associated with $p$. The regular representation of $\fS_n$,
$V$, which consists of linear combinations $\sum_{g\in\fS_n} x_g g$, of elements of $\fS_n$,
contains $d_p$ representations of the type labeled by $p$, so that $|\fS_n|=\sum_p d_p^2$, which we can regard as being labeled by the standard Young tableaux associated with $p$. We label these $\lambda_i^p, 1\leq i\leq d_p$.

Given a Young tableau, $\lambda$, we define  $\fA_\lambda$ of $\fS_n$ to be the the subgroup of $\fS_n$ consisting those 
permutations which map each row of $\lambda$ into itself and define $\fB_\lambda$ of $\fS_n$ to be the the subgroup of $\fS_n$ consisting those 
permutations which map each column of $\lambda$ into itself. Let
\be a_\lambda=\sum_{\varrho\in\fA_\lambda}\varrho,\qquad
b_\lambda=\sum_{\varrho\in\fB_\lambda}\epsilon(\varrho)\varrho,\label{defab}\ee
where $\epsilon(\varrho)$ denotes the sign of the permutation $\varrho$.  Then
$$\varrho a_\lambda=a_\lambda\varrho=a_\lambda,\quad \varrho\in\fA_\lambda;\qquad
\varrho b_\lambda=b_\lambda\varrho=\epsilon(\varrho)b_\lambda,\quad \varrho\in\fB_\lambda.$$
Define the Young symmetrizer
\be c_\lambda=a_\lambda b_\lambda.\label{Youngsym}\ee
Then 
\be c_\lambda^2=N_pc_\lambda,\qquad\hbox{where } N_p={n!\over d_\lambda},\label{csq}\ee
and 
\be c_\lambda c_\mu=0,\label{cortho}\ee
if $\lambda,\mu$ have different shapes, {\it i.e.} are associated with different Young diagrams (partitions). 

If the distinct Young tableaux   $\lambda,\mu$ are associated with the same Young diagram, $p$, we can find a permutation $\sigma_{\lambda\mu}\in\fS_n$, which takes
$\mu$ into $\lambda$; then
\be a_\lambda=\sigma_{\lambda\mu}a_\mu\sigma_{\mu\lambda},\qquad
b_\lambda=\sigma_{\lambda\mu}b_\mu\sigma_{\mu\lambda},\qquad
c_\lambda=\sigma_{\lambda\mu}c_\mu\sigma_{\mu\lambda},\qquad 
\sigma_{\mu\lambda}=\sigma_{\lambda\mu}^{-1}.\label{sigmalamdamu}\ee
Further (see \cite{GW}, page 393, or \cite{DEL}, page 75),
{\it either} there exists a pair $(j,k)$ contained in a single column of $\lambda$ and a single column of $\mu$, in which case, if $t\in\fS_n$ is the transposition interchanging $j$ and $k$, $t\in\fB_\lambda\cap\fA_\mu$, $t^2=1$, so that
\be b_\lambda a_\mu=b_\lambda t^2 a_\mu=-b_\lambda a_\mu,\qquad\hbox{implying} \quad c_\lambda c_\mu=0,\label{sigalt1}\ee
{\it or} the elements of each given column of $\lambda$ are in different rows in $\mu$, in which case 
\be\sigma_{\lambda\mu}=\beta_{\lambda\mu}\alpha_{\lambda\mu},\qquad\hbox{for some }
\alpha_{\lambda\mu}\in\fA_\mu,\beta_{\lambda\mu}\in\fB_\lambda,\label{sba}\ee
so that
\be b_\lambda a_\mu=\epsilon_{\lambda\mu}b_\lambda \beta_{\lambda\mu}\alpha_{\lambda\mu}a_\mu
=\epsilon_{\lambda\mu}b_\lambda \sigma_{\lambda\mu}a_\mu,\qquad\hbox{where }\epsilon_{\lambda\mu}=\epsilon(\beta_{\lambda\mu}),
\label{sigalt2}\ee
implying
$$c_\lambda c_\mu=N_p\epsilon_{\lambda\mu}\sigma_{\lambda\mu}c_\mu=N_p\epsilon_{\lambda\mu}c_\lambda\sigma_{\lambda\mu}.$$
If the normalized Young tableau $\hat c_\lambda=c_\lambda/N_p$, and $\hat a_\lambda =a_\lambda/\sqrt N_p, 
\hat b_\lambda =b_\lambda/\sqrt N_p,$ and $\lambda, \mu$ have the same shape,
\be\hat b_\lambda \hat a_\mu= \epsilon_{\lambda\mu}\hat b_\lambda\sigma_{\lambda\mu}\hat a_\mu,
\qquad \hat c_\lambda \hat c_\mu=\epsilon_{\lambda\mu}\hat c_\lambda\sigma_{\lambda\mu}
=\epsilon_{\lambda\mu}\sigma_{\lambda\mu}\hat c_\mu,\label{bacc}\ee
where 
$\epsilon_{\lambda\lambda}=1$, $\epsilon_{\lambda\mu}=\epsilon(\beta_{\lambda\mu})$ if the elements of each given column of $\lambda$ are in different rows in $\mu$, and $\epsilon_{\lambda\mu}=0$ otherwise.

Writing,  $\lambda_i\equiv\lambda_i^p, 1\leq i\leq d_p$ for the $d_p$ standard Young tableaux of type $p$, in lexicographical order, 
that is if $i<j$ and we compare the entries of integers in the boxes of $\lambda_i$ and $\lambda_j$ reading along each row from left to right starting with the first row and proceeding to the second, and so on, then for the first discrepancy the integer in the relevant box in $\lambda_j$ is greater than the one in the corresponding box in $\lambda_i$; in this case we write $\lambda_i <\lambda_j$ if $i<j$.
Then, writing $a_i=a_{\lambda_i}, b_i=b_{\lambda_i}, \sigma_{ij}=\sigma_{\lambda_i\lambda_j}$,  $b_ia_j=0$ if $i>j$, 
\be\hat b_i \hat a_j= \epsilon_{ij}\hat b_i\sigma_{ij}\hat a_j,\label{baab}\ee
where $\epsilon_{ij}$ is defined as in \eqref{bacc}, for $i\leq j$.

For each Young tableau $\lambda$ of type $p$, 
\be V_\lambda=Vc_\lambda,\label{defVlambda}\ee
defines an irreducible representation subspace of the regular representation $V$ of type $p$, dimension $d_p$. The spaces
$V_{\lambda_i}$, $1\leq i\leq d_p$, provide $d_p$ irreducible representations of type $p$ in $V$. In fact,
\be V\cong \bigoplus_p\bigoplus_{i=1}^{d_p} V_{\lambda^p_i}.\label{Vsum}\ee
Corresponding to this decomposition into irreducible components, a basis for $V$ is provided by
\be\{\sigma_{\lambda^p_i\lambda^p_j}\hat c_{\lambda^p_j}=\hat a_{\lambda^p_i}\sigma_{\lambda^p_i\lambda^p_j}\hat b_{\lambda^p_j}: 1\leq i,j\leq d_p; p\in P(n)\}\label{basV}\ee
where $P(n)$ denotes the set of partitions of $n$. To establish that this is a basis, it is enough to show that the states
$$\hat  a_i\sigma_{ij}\hat b_j=\sigma_{ij}\hat c_j,\qquad 1\leq i,j\leq d_p,$$
using the notation of \eqref{baab}, are linearly independent. If 
$$\sum_{1\leq i,j\leq d_p} x_{ij}\sigma_{ij}\hat c_j=0,\quad\hbox{then}\quad
\sum_{1\leq i,j\leq d_p} x_{ij}\hat c_\ell\sigma_{ij}\hat c_j\hat c_k=0,$$
implying
\be\sum_{\ell\leq i,j\leq k} x_{ij}\hat c_\ell\sigma_{ij}\hat c_j\hat c_k=0.\label{xcscc}\ee
Suppose some $x_{ij}\ne 0$; choose $\ell$ so that is the largest value of $i$ for which this is true and then $k$ so that it is the smallest value of $j$ for which $x_{\ell j}\ne0$. Then all the terms on the left hand side of \eqref{xcscc} are zero except for one leaving
$$x_{\ell k}\hat c_\ell\sigma_{\ell k}\hat c_k\hat c_k=x_{\ell k}\hat c_\ell\sigma_{\ell k}=0,$$
which implies $x_{\ell, k}=0$, a contradiction. Thus, we conclude that $x_{ij}=0$ for all $i,j$ and so that the states  \eqref{basV} form a basis.

Now we seek to determine $x_{ij}, 1\leq i < j<d_p$, so that 
\be P_p=\sum_{1\leq i\leq d_p}\hat c_i+\sum_{1\leq i<j\leq d_p}x_{ij}\sigma_{ij}\hat c_j\label{defPp}\ee
is the projection on to the spaces corresponding to the standard Young tableaux of shape $p$,
\be V_p=\bigoplus_{i=1}^{d_p} V_{\lambda^p_i}.\label{Vpsum}\ee
(See \cite{DEL}, page 76.) A necessary and sufficient condition for this is $P_p\sigma_{ij}\hat c_j= \sigma_{ij}\hat c_j$ for $1\leq i,j\leq d_p$. If this holds we will have $\sum_p P_p=1$, because $P_{p'}\sigma_{ij}\hat c_j=0$ if $p'$ is another shape of Young tableaux.
\begin{align}
P_p\sigma_{k\ell}\hat c_\ell&=\sum_{1\leq i\leq d_p}\hat c_i\hat c_k\sigma_{k\ell}+\sum_{1\leq i<j\leq d_p}x_{ij}\sigma_{ij}\hat c_j\hat c_k\sigma_{k\ell}\cr
&=\hat c_k^2\sigma_{k\ell}+\sum_{1\leq i<k}\hat c_i\hat c_k\sigma_{k\ell}+
\sum_{1\leq i<k}x_{ik}\sigma_{ik}\hat c_k^2\sigma_{k\ell}
+\sum_{1\leq i<j<k}x_{ij}\sigma_{ij}\hat c_j\hat c_k\sigma_{k\ell}\cr
&=\sigma_{k\ell}\hat c_\ell
+\sum_{1\leq i<k}\epsilon_{ik}\sigma_{i\ell}\hat c_\ell^2+
\sum_{1\leq i<k}x_{ik}\sigma_{i\ell}\hat c_\ell
+\sum_{1\leq i<j<k}x_{ij}\epsilon_{jk}\sigma_{i\ell}\hat c_\ell^2\cr
&=\sigma_{k\ell}\hat c_\ell
+\sum_{1\leq i<k}\left(\epsilon_{ik}+x_{ik}
+\sum_{ i<j<k}x_{ij}\epsilon_{jk}\right)\sigma_{i\ell}\hat c_\ell.
\end{align}
Because the  $\sigma_{i\ell}\hat c_\ell$ are linearly independent, the condition that $P_p\sigma_{k\ell}\hat c_\ell= \sigma_{k\ell}\hat c_\ell$ is 
\be x_{ik}=-\epsilon_{ik}
+\sum_{ i<j<k}x_{ij}\epsilon_{jk},\qquad\hbox{for } 1\leq i<k.\label{xeqn}\ee
We can solve this equation iteratively for increasing $k-i$, starting with $k-i=1$:
\begin{align}
x_{k-1,k}&=-\epsilon_{k-1,k};\cr
x_{k-2,k}&=-\epsilon_{k-2,k}+\epsilon_{k-2,k-1}\epsilon_{k-1,k};\cr
&\hskip-10truemm\ldots\,\, \ldots\cr
x_{ik}&=-\epsilon_{ik}
+\sum_{ i<j<k}\epsilon_{ij}\epsilon_{jk}-\sum_{ i<j<\ell<k}\epsilon_{ij}\epsilon_{j\ell}\epsilon_{\ell k}
+\ldots+(-1)^{k-i}\epsilon_{i,i+1}\epsilon_{i+1,i+2}\ldots\epsilon_{k-1, k},\cr
\end{align}
for $i<k$. Then 
\be P_p={d_p\over n!}\sum_{1\leq j\leq d_p}\xi_j c_j, \qquad\hbox{where}\quad\xi_j= 1+\sum_{1\leq i<j}x_{ij}\sigma_{ij}\label{Ppeqn}\ee
and
\be 1={1\over n!}\sum_pd_p
\sum_{1\leq i\leq d_p} \xi_{\lambda_i^p}c_{\lambda_i^p}.\label{resunity}\ee
In fact $\xi_j=1$ for all $j$ when $n\leq 4$.

Every $\fB_\lambda\ne 1$ unless $\lambda$ corresponds to the Young diagram $p_1$ with only one row; this corresponds to the identity representation and has $\fA_\lambda=\fS_n$. So
\be 1={1\over n!}\sum_{\varrho\in\fS_n}\varrho
+{1\over n!}\sum_{p\ne p_1}d_p
\sum_{1\leq i\leq d_p} \xi_{\lambda_i^p}c_{\lambda_i^p}.\label{resunity2}\ee
So, for given $\kappa_{n-1}\in\K_{n-1}$, if $\kappa_n$ is the solution to \eqref{kappaeq4} and \eqref{kcycle4} orthogonal to all symmetric invariant tensors, 
\be \kappa_n={1\over n!}\sum_{p\ne p_1}d_p
\sum_{1\leq i\leq d_p} \xi_{\lambda_i^p}c_{\lambda_i^p}\kappa_n.\label{kappares}\ee

For each Young tableau $\lambda$ not corresponding to the identity representation, choose a transposition $t_\lambda\in\fB_\lambda$. Then $c_\lambda t_\lambda = -c_\lambda$ so that 
\be c_\lambda\kappa_n=\half c_\lambda(1-t_\lambda)\kappa_n=\half c_\lambda\phi(t_\lambda,\kappa_{n-1})\ee
and 
\be \kappa_n={1\over 2n!}\sum_{p\ne p_1}d_p
\sum_{1\leq i\leq d_p} \xi_{\lambda_i^p}c_{\lambda_i^p}
\phi(t_{\lambda_i^p},\kappa_{n-1}).\label{kappares2}\ee
We can express each of the transpositions $t_\lambda$ as a product of the generating transpositions $\sigma_i$, defined as in \eqref{sigmai}; if $t_\lambda$ is the transposition interchanging $j$ and $k$, with $j<k$,
$$ t_\lambda=\sigma_{k-1}\ldots\sigma_{j+1}\sigma_j\sigma_{j+1}\ldots\sigma_{k-1},$$
We can then evaluate $\phi(t_\lambda,\kappa_{n-1})$ for $\lambda=\lambda_i^p$, using \eqref{phirhorho2}, to give an explicit expression for $\kappa_n$ in terms of $\kappa_{n-1}$.
\vfil\eject

\section{Properties of $H_n$}
\label{sect:PropHn}

In this Appendix we establish some properties of the generating function $H_n$, defined by
\eqref{Hnform}. We noted that 
\be
{1\over\nu^n}H_n(\nu_{12},\ldots,\nu_{n1},\tau;\nu)={1\over\sigma(\nu,\tau)^n}\prod_{j=1}^n{\sigma(\mu_j+\nu,\tau)\over\sigma(\mu_j,\tau)}
\label{lab134}\ee
is an elliptic function of $\nu$ and $\nu_j$, $1\leq j\leq n$. Viewed as function of $\nu$, it has a pole of order $n$ at the origin but it is otherwise regular. 
\be{1\over\nu^n}H_n={1\over\nu^n}+{1\over\nu^{n-1}}H_{n,1}+\ldots +
{1\over \nu^2}H_{n,n-2}+{1\over \nu}H_{n,n-1}+H_{n,n}+{\cal O}(\nu)\qquad\hbox{as }\nu\rightarrow 0.
\label{lab135}\ee
Writing the Weierstrass elliptic function
\be\P(\nu,\tau)=\nu^{-2}+\sum_{m=1}^\infty c_{2m}(\tau)\nu^{2m},\label{lab136}\ee
we note that its derivatives 
\be\P^{(n)}(\nu,\tau)
={(-1)^n(n+1)!\over\,\nu^{n+2}}+
n! \,c_n(\tau)+{\cal O}(\nu),\label{lab137}\ee
as $\nu\rightarrow 0$, where $c_\ell=0$ if $\ell$ is odd. So
\be
{1\over\nu^n}H_n-{(-1)^n\over (n-1)!}\P^{(n-2)}(\nu)+{(-1)^n\over (n-2)!}\P^{(n-3)}(\nu)H_{n,1}+\ldots
-\P(\nu)H_{n,n-2}\label{expHn}\ee
is an elliptic function of $\nu$ whose only potential singularity is a simple pole at the origin. This implies that it is constant as a function of $\nu$ (see \cite{RM}, Proposition 4.11, page 48)
and the residue of the pole must vanish:
\be H_{n,n-1}(\nu_{12},\ldots,\nu_{n1},\tau)=0,\qquad\hbox{for } n\geq 2.\label{lab139}\ee
For $n=3$, this gives
\be \left[\sum_{r=1}^3\zeta(\mu_r)\right]^2=\sum_{r=1}^3\P(\mu_r),
\qquad\hbox{if }\sum_{r=1}^3\mu_r=0;\ee
and, for $n=4$,
\be \left[\sum_{r=1}^4\zeta(\mu_r)\right]^3
= 3\left[\sum_{r=1}^4\zeta(\mu_r)\right]\left[\sum_{r=1}^4\P(\mu_r)\right]
+\left[\sum_{r=1}^4\P'(\mu_r)\right],
\qquad\hbox{if }\sum_{r=1}^4\mu_r=0;\ee
(see \cite{WW}, pages 446 and 459, respectively).

We can equate \eqref{expHn} to its value at $\nu=0$, giving
\begin{align}
{1\over\nu^n}H_n={(-1)^n\over (n-1)!}\P^{(n-2)}(&\nu)-{(-1)^n\over (n-2)!}\P^{(n-3)}(\nu)H_{n,1}\cr
&+\ldots
+\P(\nu)H_{n,n-2}+H_{n,n}
-\sum_{\ell=1}^{[\half n]-1}2\ell c_{2\ell} H_{n,n-2\ell-2}.\label{lab140}\end{align}
By direct calculation  for $n=2$, we have that
\be\nu^{-2}H_2(\nu_{12},\nu_{21},\tau;\nu)={\sigma(\nu+\nu_{12})\sigma(\nu+\nu_{21})
\over\sigma(\nu)^2\sigma(\nu_{12})\sigma(\nu_{21})}
=\P(\nu)-\P(\nu_{12}),\label{lab142}\ee
so that
\be H_{2,0}(\nu_{12},\tau)=1;\qquad H_{2,2}(\nu_{12},\tau)=-\P(\nu_{12},\tau);\ee
\be H_{2,2m}(\nu_{12},\tau)=c_{2m-2}(\tau), \,\,m\geq 2.\label{lab143}\ee
 In particular, from the expression given for $H_{2,4}$ given by \eqref{explicitH}, we can deduce the differential equation for $\P$,
$$\P''-6\P^2+\half g_2(\tau)=0,$$
where
$$g_2(\tau)={\scriptstyle {2\over 3}}\pi^4\left(\theta_2(0,\tau)^8 
+\theta_3(0,\tau)^8+\theta_4(0,\tau)^8\right) = 20 c_2(\tau).$$

Consider the symmetrizations of $H_{n,m}$ and $H_n$,
\be H_{n,m}^{S}(\nu_{12},\ldots,\nu_{n1},\tau)=
{1\over n!}\sum_{\varrho\in\fS_n}H_{n,m}(\nu_{\varrho(1)\varrho(2)},\ldots,\nu_{\varrho(n)\varrho(1)},\tau),\label{lab145a}\ee
\be H_n^{S}(\nu_{12},\ldots,\nu_{n1},\tau;\nu)=
{1\over n!}\sum_{\varrho\in\fS_n}H_{n}(\nu_{\varrho(1)\varrho(2)},\ldots,\nu_{\varrho(n)\varrho(1)},\tau;\nu).\label{lab145b}\ee
Since $ H_n(\nu_{12},\ldots,\nu_{n1},\tau;\nu)=H_n(-\nu_{1n},\ldots,-\nu_{21},\tau;\nu),$
$H_n^{S}$ is an even function of $\nu$ and so
\be H_{n,m}^{S}(\nu_{12},\ldots,\nu_{n1},\tau)=0,\,\,m \hbox{ odd}.
\label{lab146}\ee
For $n>2$,
\be \Res{\nu_{12}}
H_{n,m}^S(\nu_{12},\ldots,\nu_{n1},\tau)=0;\qquad
\Res{\nu_{12}}
H_n^S(\nu_{12},\ldots,\nu_{n1},\tau;\nu)=0.\label{lab147}\ee
Because $\nu^{-n}H_{n,m}^{S}(\nu_{12},\ldots,\nu_{n1},\tau)$ and $\nu^{-n}H_n^{S}(\nu_{12},\ldots,\nu_{n1},\tau;\nu)$
are elliptic as functions of $\nu_{ij}$, but have no poles in these variables, it follows that they are independent of them, {\it i.e.}
\be H_{n,m}^{S}(\nu_{12},\ldots,\nu_{n1},\tau)=H_{n,m}^{S}(\tau),\qquad\hbox{independent of $\nu_{ij}$};\ee
\be H_n^{S}(\nu_{12},\ldots,\nu_{n1},\tau;\nu)=H_n^{S}(\tau;\nu)
=\sum_{m=0}^\infty H_{n,m}^{S}(\tau)\nu^m,\qquad n>2.\label{lab148}\ee
For $n=2$, we have that
\be H_2^S(\nu_{12},\nu_{21},\tau;\nu)=H_2(\nu_{12},\nu_{21},\tau;\nu)
=\nu^2[\P(\nu)-\P(\nu_{12})]\label{lab149}\ee
So, 
\be H^S_{2,0}(\nu_{12},\tau)=1;\qquad H^S_{2,2}(\nu_{12},\tau)=-\P(\nu_{12},\tau);\ee
\be H^S_{2,2m}(\nu_{12},\tau)=c_{2m-2}(\tau), \,\,m\geq 2.\label{lab150}\ee

By direct calculation  for $n=3$, we have that
\begin{align}
H_3^S(\nu_{12},&\nu_{23},\nu_{31},\tau;\nu)\cr
&={\nu^3\sigma(\nu+\nu_{12})\sigma(\nu+\nu_{23})\sigma(\nu+\nu_{31})
\over2\sigma(\nu)^3\sigma(\nu_{12})\sigma(\nu_{23})\sigma(\nu_{31})}
+{\nu^3\sigma(\nu+\nu_{21})\sigma(\nu+\nu_{13})\sigma(\nu+\nu_{32})
\over2\sigma(\nu)^3\sigma(\nu_{21})\sigma(\nu_{13})\sigma(\nu_{32})}\cr
&=-\half\nu^3\P'(\nu)=\half\nu^3\zeta''(\nu).\label{H3S}\end{align}

We also have, by direct calculation, that
\be H^S_n(\tau;\nu)={(-1)^n\over (n-1)!}\nu^n\P^{(n-2)}(\nu)={(-1)^{n+1}\over (n-1)!}\nu^n\zeta^{(n-3)}(\nu) \qquad\hbox{ holds for }3\leq n\leq 6.
\label{conjecture}\ee
and conjecture that it holds for all $n\geq 3$. 
When \eqref{conjecture} holds, we have that
\be
H^S_{n,0}(\tau)=1;\qquad H^S_{n,2m}(\tau)=0,\,\,1\leq m<\half n;\label{lab00}\ee
\be H^S_{n,2m}(\tau)={(-1)^n(2m-2)!\over (n-1)!(2m-n)!}c_{2m-2}(\tau),\,\,m\geq \half n.\label{lab156}\ee
\vfil\eject
\section{Explicit Formulae for Two-, Three- and Four-Point Loops}
\label{sect:formulae}

In this appendix we show how the formulae we have given previously \cite{DG} for two-, three- and four-point current algebra loops relate to the general result \eqref{general}. For two- and three-point loops there is no distinction between the connected part and the whole loop amplitude.

\subsection{\sl Two-point Current Algebra Loop}

\begin{align}\tr \left(J^a(\rho_1)J^b(\rho_2)w^{L_0}\right)
&= -{\chi(\tau)\over4\pi^2\rho_1\rho_2}\delta^{ab} k \chi(\tau)\,
\left[\left(\chi_{NS}^{12}\right)^2 -4\pi^2 f(\tau)\right]\cr
&= { \chi(\tau) \over4\pi^2\rho_1\rho_2}
\left(\kappa_2^{ab} H_{2,2} + \kappa_{2,0}^{ab}\right)
\end{align}
where
\begin{align}
 H_{2,2}
 &= -{\cal P}_{12} =- \left(\chi_{NS}^{12}\right)^2 - {\pi^2\over 3} \left[\theta_2^4(0,\tau)-\theta_4^4(0,\tau)\right]\\
 \kappa_2^{ab} &=k \delta^{ab},
\quad\hbox{for }\,\kappa^{ab}=k\delta^{ab},\\
\kappa_{2,0}^{ab}   
&= {4\pi^2\over\chi(\tau)} \,\tr(J^a_0J^b_0 w^{L_0}) \,\, +k\delta^{ab} 
{1\over 3} {\theta_1'''(0,\tau)\over\theta_1'(0,\tau)}\cr
&= 4\pi^2k \delta^{ab} \left[ f(\tau) + \twelfth
(\theta_2^4(0,\tau)-\theta_4^4(0,\tau) \right]
\end{align}
and
\be
f(\tau) = {\chi^{(2)}(\tau)\over k\chi(\tau)} + {1\over 4\pi^2}
{\theta_3''(0,\tau)\over\theta_3(0,\tau)} \,,
\qquad \tr(J^a_0J^b_0 w^{L_0}) = \delta^{ab} \chi^{(2)}(\tau),\ee
\be\chi_{NS}^{ij}=\chi_{NS}(\nu_i-\nu_j,\tau), \qquad\chi(\tau)=\tr\left(w^{L_0}\right).\ee

\subsection{\sl Three-point  Current Algebra Loop}

\begin{align}\tr\left(J^a(\rho_1)J^b(\rho_2)J^c(\rho_3)w^{L_0}\right)
\hskip-35truemm&\cr
&=-{i \chi(\tau) \over8\pi^3\rho_1\rho_2\rho_3} 
\left[ \kappa_3^{abc} H_{3,3} + \kappa_{3,1}^{abc} H_{3,1} + \kappa_{3,0}^{abc}
\right]_\Sop\cr
&={ ik f^{abc}\chi(\tau)\over8\pi^3\rho_1\rho_2\rho_3} \left[ \chi_{NS}^{21}\chi_{NS}^{32}
\chi_{NS}^{13} -4\pi^2(\zeta^{21}+\zeta^{32}+\zeta^{13})f(\tau)\right]
+ {d^{abc} \chi^{(3)}(\tau)\over2\rho_1\rho_2\rho_3}
\end{align}where  $\zeta^{ij} = \zeta(\nu_j-\nu_i, \tau)$  and
$\Phi_\Sop$ denotes  the symmetrization 
\be\Phi^{a_{i_1}a_{i_2}\ldots a_{i_n}}_\Sop(\nu_{i_1},\nu_{i_2},\ldots,\nu_{i_n},\tau)
={1\over n}\sum_{\varrho\in\fS_n}\A^{a_{i_{\varrho(1)}}a_{i_{\varrho(2)}}\ldots a_{i_{\varrho(n)}}}(\nu_{i_{\varrho(1)}},\nu_{i_{\varrho(2)}},\ldots,\nu_{i_{\varrho(n)}},\tau);\label{Sop}
\ee
and
\begin{align}H_{3,3} &= {1\over 6} \left[ (\zeta^{21} + \zeta^{32} + \zeta^{13})^2- 3 ( \zeta^{21} + \zeta^{32} + \zeta^{13}) ({\cal P}^{21} + {\cal P}^{32} + {\cal P}^{13})
-  ({\cal P'}^{21} + {\cal P'}^{32} + {\cal P'}^{13})\right]\cr
&=  {1\over 6} \left[- 2( \zeta^{21} + \zeta^{32} + \zeta^{13})
({\cal P}^{21} + {\cal P}^{32} + {\cal P}^{13})
-  ({\cal P'}^{21} + {\cal P'}^{32} + {\cal P'}^{13})\right]\hskip10truemm
{\rm since\,} H_{3,2} = 0\cr
&=- \chi_{NS}^{21}\chi_{NS}^{32}\chi_{NS}^{13}
\,\,- {\pi^2\over 3} (\theta_2^4(0,\tau) -\theta_4^4(0,\tau))\,(\zeta^{21} + \zeta^{32}
+\zeta^{13})\\
H_{3,1} &= \zeta^{21}+\zeta^{32}+\zeta^{13}\\
 \kappa_3^{abc} &=\half k f^{abc}  \\
 \kappa_{3,1}^{abc} &= 2\pi^2  k f^{abc} 
 \left [ f(\tau) +\twelfth \left(\theta_2^4(0,\tau) - \theta_4^4(0,\tau)\right)\right]\\
 \kappa_{3,0}^{abc} 
&= 2\pi^3 i d^{abc}{\chi^{(3)}(\tau) \over\chi(\tau)}
={ 4\pi^3i\over\chi(\tau)}\tr \left(J^a_0J^b_0J^c_0 w^{L_0}\right)_\Sop\end{align}
\be\tr (J^a_0J^b_0J^c_0 w^{L_0}) = \half f^{abc} \chi^{(2)}(\tau)
+\half  d^{abc} \chi^{(3)}(\tau)\ee
and we also use \hskip2truemm$\kappa_{3,m}^{abc} - \kappa_{3,m}^{bac} = {f^{ab}}_e \kappa_{2,m-1}^{ec},\quad
 \kappa_{3,m}^{abc} = \kappa_{3,m}^{cab},$ \hskip2truemm where $\kappa_{n,n}=\kappa_n$.
\subsection{\sl Four-point Current Algebra Loop}
In \cite{DG}, we gave the general form of the four-point loop in the symmetric form
\begin{align}
\tr(J^a(\rho_1)&J^b(\rho_2) J^c(\rho_3) J^d(\rho_4)w^{L_0})\rho_1\rho_2\rho_3
\rho_4\cr
= &
\delta^{ab}\delta^{cd}\, \left (k^2 \chi(\tau)\,
[(\chi_{NS}^{12})^2/4\pi^2-  f(\tau)] [(\chi_{NS}^{34})^2/4\pi^2- f(\tau)]-\chi^{(2)}(\tau)^2/
\chi(\tau)\right)\cr
&+ \delta^{ac}\delta^{bd}
\, \left( k^2 \chi(\tau)\,
[(\chi_{NS}^{13})^2/4\pi^2-  f(\tau)] [(\chi_{NS}^{24})^2/4\pi^2-  f(\tau)]-\chi^{(2)}(\tau)^2/
\chi(\tau)\right)\cr
&\hskip0pt + \delta^{ad}\delta^{bc}
\, \left(k^2 \chi(\tau)\,
[(\chi_{NS}^{14})^2/4\pi^2-  f(\tau)] [(\chi_{NS}^{23})^2/4\pi^2-  f(\tau)]-\chi^{(2)}(\tau)^2
/\chi(\tau)\right)\cr
& + \tr (J^a_0 J^b_0 J^c_0J^d_0 w^{L_0})_S\cr
&- {1\over 96}\left (\sigma^{abcd} + \sigma^{adcb} + \sigma^{acdb}
+ \sigma^{abdc} +
\sigma^{adbc} +  \sigma^{acbd}\right)
\chi(\tau)\theta_2^4(0,\tau)\theta_4^4(0,\tau)\cr
& - \left (\sigma^{abcd}
+ \sigma^{adcb}
\right){1\over 32\pi^4} \chi(\tau)\,\left\{
\chi_{NS}^{12}\chi_{NS}^{23}\chi_{NS}^{34}\chi_{NS}^{41} \phantom{{{\cal P}'_{24}
- {\cal P}'_{14}\over
{\cal P}_{24}  - {\cal P}_{14}}}\right.\cr
&\hskip15truemm\left.- \pi^2 f(\tau)
\left({{\cal P}'_{24} - {\cal P}'_{32} \over {\cal P}_{24}
- {\cal P}_{32}}\right)
\left( {{\cal P}'_{24}  - {\cal P}'_{41}\over{\cal P}_{24}
- {\cal P}_{41}}\right )
+  4\pi^2f(\tau)\P_{24}\right\}\cr
& - \left(\sigma^{acdb}
+ \sigma^{abdc} \right){1\over 32\pi^4} \chi(\tau)\,\left\{
\chi_{NS}^{13}\chi_{NS}^{34}\chi_{NS}^{42}\chi_{NS}^{21} \phantom{{{\cal P}'_{24}
- {\cal P}'_{14}\over
{\cal P}_{24}  - {\cal P}_{14}}}\right.\cr
&\hskip15truemm\left.- \pi^2 f(\tau)
\left({{\cal P}'_{24}  - {\cal P}_{32}\over{\cal P}_{24} - {\cal P}_{32}}\right)
\left({{\cal P}'_{21} - {\cal P}'_{32}\over{\cal P}_{21} - {\cal P}_{32}}\right)
+  4\pi^2 f(\tau) {\cal P}_{32}\right\}\cr
& -  \left(  \sigma^{adbc} +  \sigma^{acbd}\right)\,\,{1\over 32\pi^4} \chi(\tau)\left\{
\chi_{NS}^{14}\chi_{NS}^{42}\chi_{NS}^{23}\chi_{NS}^{31}\phantom{{{\cal P}'_{24}
- {\cal P}'_{14}\over
{\cal P}_{24}  - {\cal P}_{14}}}\right.\cr
&\hskip15truemm\left. - \pi^2 f(\tau) \left (
{{\cal P}'_{24}  - {\cal P}'_{32}\over {\cal P}_{24}  - {\cal P}_{32}}\right )
\left( {{\cal P}'_{14} - {\cal P}'_{31}\over
{\cal P}_{14} - {\cal P}_{31}}\right)  + 4\pi^2 f(\tau)
{\cal{P}}_{34}\right\}\cr
&- {i\over 4\pi}\chi^{(3)}(\tau)\left[\sigma^{abcd} \left(\zeta^{21}
+ \zeta^{32} + \zeta^{43}+\zeta^{14}\right)
+ \sigma^{adcb}\left(\zeta^{41} + \zeta^{34} + \zeta^{23}+\zeta^{12}\right)
\right.\cr
&\hskip21truemm+ \sigma^{acdb} \left(\zeta^{31} + \zeta^{43} + \zeta^{24}+\zeta^{12}\right)
+ \sigma^{abdc} \left(\zeta^{21} + \zeta^{42} + \zeta^{34}+\zeta^{13}\right)\cr
&\left.\hskip21truemm+ \,\sigma^{adbc}\left(\zeta^{41} + \zeta^{24}
+ \zeta^{32}+\zeta^{13}\right)
 +  \sigma^{acbd}
\left(\zeta^{31} + \zeta^{23} + \zeta^{42}+\zeta^{14}\right)\right],\cr
\end{align}
where $\P_{ij}=\P(\nu_j-\nu_i,\tau)$, $\P'_{ij}=\P'(\nu_j-\nu_i,\tau)$, and
$\sigma^{abcd}=\tr(t^at^bt^ct^d)$, where $t^a$ provides a representation of 
$\g$ with 
$\sigma^{ab}=\tr(t^at^b)= 2 \kappa^{ab}=2 k\delta^{ab}$ 
and $\sigma^{abc}
=\tr(t^at^bt^c)={f^{ab}}_e\kappa^{ec}+d^{abc};$ then
\be
\sigma^{abcd} ={k\over 3}\left(  f^{ab}_{\hskip 6pt e} f^{cde}
+ f^{da}_{\hskip 6pt e} f^{bce}\right)+{1\over 4}
\left(  f^{ab}_{\hskip 6pt e} d^{ecd}+f^{ac}_{\hskip 6pt e} d^{edb}
+ f^{bc}_{\hskip 6pt e} d^{eab}\right) +  \bar \omega^{abcd}
\ee
where  $\bar \omega^{abcd}$ is a totally symmetric tensor, independent
of $\tau$.
Now, using 
\begin{align}
{1\over 2}&\left({{\cal P}'_{24}  - {\cal P}_{32}\over{\cal P}_{24} - {\cal P}_{32}}\right)
\left({{\cal P}'_{24} - {\cal P}'_{41}\over{\cal P}_{24} - {\cal P}_{41}}\right)
-  2  {\cal P}_{24}=\cr
&=\left(\zeta^{13}+ \zeta^{32} + \zeta^{21}\right)
\left(\zeta^{13}+ \zeta^{34} + \zeta^{41}\right)+
\left(\zeta^{24}+ \zeta^{41} + \zeta^{12}\right)
\left(\zeta^{24}+ \zeta^{43} + \zeta^{32}\right)-{\cal P}_{13} - {\cal P}_{24}\cr
&={\cal P}_{12}+{\cal P}_{23}+{\cal P}_{34}+{\cal P}_{41}
-(\zeta^{12}+\zeta^{23}+\zeta^{34}
+\zeta^{41})^2,
\end{align}
we have that
\begin{align}
\tr(&J^a(\rho_1)J^b(\rho_2)J^c(\rho_3)J^d(\rho_4)w^{L_0})_C\cr
&=\tr(J^a(\rho_1)J^b(\rho_2) J^c(\rho_3) J^d(\rho_4)w^{L_0})
-\tr(J^a(\rho_1)J^b(\rho_2)w^{L_0})\tr(J^c(\rho_3)J^d(\rho_4)w^{L_0})\cr
& -\tr(J^a(\rho_1)J^c(\rho_3)w^{L_0}) \,\,\tr(J^b(\rho_2)J^d(\rho_4)
w^{L_0})- \tr(J^a(\rho_1)J^d(\rho_4)w^{L_0}) \,\,\tr(J^c(\rho_3)
J^b(\rho_2)w^{L_0})\cr
&=- { \chi(\tau)\over 16\pi^4 \rho_1\rho_2\rho_3\rho_4} 
\left[\kappa_4^{abcd} H_{4,4} + \kappa_{4,2}^{abcd} H_{4,2}
+ \kappa_{4,1}^{abcd}H_{4,1}+\kappa_{4,0}^{abcd}\right]_\Sop
\end{align}
with $\Sop$ denoting symmetrization as defined in \eqref{Sop}, and
\begin{align}24 H_{4,4} &= (\zeta^{21}+\zeta^{32}+\zeta^{43}+\zeta^{14})^4
- 6 (\zeta^{21}+\zeta^{32}+\zeta^{43}+\zeta^{14})^2({\cal P}_{21}
+ {\cal P}_{32} + {\cal P}_{43} + {\cal P}_{14})\cr
&\hskip7pt- 4 (\zeta^{21}+\zeta^{32}+\zeta^{43}+\zeta^{14})({\cal P'}_{21}
+ {\cal P'}_{32} + {\cal P'}_{43} + {\cal P'}_{14})
+ 3 ({\cal P}_{21} + {\cal P}_{32} + {\cal P}_{43} + {\cal P}_{14})^2\cr
&\hskip7pt- ({\cal P''}_{21} + {\cal P''}_{32} + {\cal P''}_{43}
+ {\cal P''}_{14})\cr
&= - 3 (\zeta^{21}+\zeta^{32}+\zeta^{43}+\zeta^{14})^2({\cal P}_{21}
+ {\cal P}_{32} + {\cal P}_{43} + {\cal P}_{14})\cr
&\hskip9pt - 3 (\zeta^{21}+\zeta^{32}+\zeta^{43}+\zeta^{14})
({\cal P'}_{21} + {\cal P'}_{32} + {\cal P'}_{43} + {\cal P'}_{14})\cr
&\hskip9pt + 3 ({\cal P}_{21} + {\cal P}_{32} + {\cal P}_{43}
+ {\cal P}_{14})^2- ({\cal P''}_{21} + {\cal P''}_{32} + {\cal P''}_{43}
+ {\cal P''}_{14}),\quad {\rm since}\, H_{4,3} = 0,\cr
&= -{\pi^2\over 3} (\theta_2^4(0,\tau)-\theta_4^4(0,\tau))\, H_{4,2}
 +  \chi_{NS}^{21}\chi_{NS}^{32}\chi_{NS}^{43}\chi_{NS}^{14}\,\,
+ {\pi^4\over 3} \theta_2^4(0,\tau)\theta_4^4(0,\tau);\cr
2H_{4,2}
&= (\zeta^{21}+\zeta^{32}+\zeta^{43}+\zeta^{14})^2- ({\cal P}_{21}
+ {\cal P}_{32} + {\cal P}_{43} + {\cal P}_{14})\cr
H_{4,1} &= \zeta^{21}
+ \zeta^{32} + \zeta^{43} +\zeta^{14}\cr
 \kappa_4^{abcd}
&= \half \sigma^{abcd} =\sixth k  
\left( f^{ab}_{\hskip6pt e} f^{cde}+ f^{da}_{\hskip6pt e} f^{bce}\right)
+\mathring{\kappa}_{4,4}^{abcd}\cr
 \kappa_{4,2}^{abcd}
&= 2\pi^2 \left[ f(\tau) +  {1\over 12}(\theta_2^4(0,\tau)
-\theta_4^4(0,\tau))\right] \sigma^{abcd}\cr
&= \twothirds\pi^2 k \left( f^{ab}_{\hskip6pt e} f^{cde}
+ f^{da}_{\hskip6pt e} f^{bce}\right) \left [f(\tau) + \twelfth
(\theta_2^4(0,\tau)-\theta_4^4(0,\tau))\right]+\mathring{\kappa}_{4,2}^{abcd}\cr
\kappa_{4,1}^{abcd} 
&= i \pi^3{\chi^{(3)}(\tau) \over\chi(\tau)}\left( f^{ab}_{\hskip 6pt e} d^{ecd}
+f^{ac}_{\hskip 6pt e} d^{edb} + f^{bc}_{\hskip 6pt e} d^{eab}\right)
+\mathring{\kappa}_{4,1}^{abcd}\cr
\kappa_{4,0}^{abcd}
& = -{8\pi^4\over 3} \left[ {1\over \chi(\tau)}\,\tr (J^a_0J^b_0J^c_0J^d_0 w^{L_0})-{1\over \chi(\tau)^2}\,
\tr(J^a_0J^b_0 w^{L_0})\,\tr(J^c_0J^d_0 w^{L_0})\right]_S\cr
&=  -{8\pi^4\over 3} \left[ {1\over \chi(\tau)}\,\tr (J^a_0J^b_0J^c_0J^d_0 w^{L_0})_S-
{\chi^{(2)}(\tau)^2\over \chi(\tau)^2}(\delta^{ab}
\delta^{cd} +\delta^{ac}\delta^{bd}+\delta^{ad}\delta^{bc})\,
\right]\cr
&=  -{8\pi^4\over 3\chi(\tau)}\tr(J_0^aJ_0^bJ_0^cJ_0^d w^{L_0})_{CS}\end{align}

where $\left[\mathring{\kappa}^{abcd}_{4,m}H_{4,m}\right]_\Sop=0,$  and we have also used
$H^S_{4,2}=H^S_{4,1}=0$ 
(which follows from \eqref{lab146} and \eqref{lab00}), and 
\be
 \chi_{NS}^{12}\chi_{NS}^{23}\chi_{NS}^{34}\chi_{NS}^{41}
 + \chi_{NS}^{13}\chi_{NS}^{34}\chi_{NS}^{42}\chi_{NS}^{21}
 +\chi_{NS}^{14}\chi_{NS}^{42}\chi_{NS}^{23}\chi_{NS}^{31}
 =-\pi^4\theta_2(0,\tau^4
\theta_4(0,\tau)^4.\ee
\vfil\eject


\providecommand{\bysame}{\leavevmode\hbox to3em{\hrulefill}\thinspace}
\providecommand{\MR}{\relax\ifhmode\unskip\space\fi MR }
\providecommand{\MRhref}[2]{%
  \href{http://www.ams.org/mathscinet-getitem?mr=#1}{#2}
}
\providecommand{\href}[2]{#2}


\begin{thebibliography}{99}

\bibitem{FZ}
I.B. Frenkel and Y. Zhu, ``Vertex Operator Algebras Associated to 
Representations of Affine and Virasoro Algebras'', 
Duke Math J. {\bf 66}, 123 (1992).\\

\bibitem{Kac} V. Kac, {\sl Infinite Dimensional Lie Algebras}, Cambridge
University Press, Cambridge, 1990.\\

\bibitem{DG} L. Dolan, and P. Goddard,
``Tree and Loop Amplitudes in Open Twistor String Theory,''
  JHEP {\bf 0706}, 005 (2007) [arXiv:hep-th/0703054].\\

\bibitem{Zhu}
Y. Zhu, ``Modular Invariance of Characters of Vertex Operator
Algebras'', J. Amer. Math. Soc. {\bf 9} (1996) 237-302.\\

\bibitem{Zhutwo}
Y. Zhu, ``Global Vertex Operators on Riemann Surfaces'',
Commun.\ Math.\ Phys.\ {\bf 485}, 165 (1994).\\

\bibitem{MT}G. Mason and M. Tuite, ``Torus Chiral n-Point Functions for Free
Boson and Lattice Vertex Operator Algebras'', Comm. Math. Phys. {\bf 235} (2003)
47-68.\\

\bibitem{MM}
S.~D.~Mathur and S.~Mukhi, 
``Correlation Functions of Current Algebra Theories on the Torus,'' Phys.\ Lett.\ B {\bf 210}, 133 (1988).\\

\bibitem{EO}
T. Eguchi and H. Ooguri, ``Conformal and Current Algebras on a General Riemann Surface'',
Nucl.\ Phys.\ B {\bf 282}, 308 (1987).\\

\bibitem{DB}
D.~Bernard, ``On The Wess-Zumino-Witten Models On The Torus,'' 
Nucl.\ Phys.\ B {\bf 303}, 77 (1988).\\

\bibitem{JNS}
T.~Jayaraman, K.~S.~Narain and M.~H.~Sarmadi, 
``SU(2)-k WZW AND Z-k Parafermion Models on the Torus,''
 Nucl.\ Phys.\ B {\bf 343}, 418 (1990).\\

\bibitem{MMS}S.D. Mathur, S. Mukhi, and A. Sen,``Differential Equations For 
Correlators And Characters In Arbitrary Rational Conformal Field Theories,'' 
Nucl.\ Phys.\ B {\bf 312}, 15 (1989).\\

\bibitem{MMStwo}
S.D. Mathur, S. Mukhi and A. Sen, ``Reconstruction of Conformal Field 
Theories from Modular Geometry on the Torus,'' Nucl.\ Phys.\ B {\bf 318}, 
483 (1989).\\

\bibitem{MMSthree}
S.D. Mathur, S. Mukhi and A. Sen, ``Correlators of Primary Fields in the 
SU(2) W Z W Theory on Riemann Surfaces,'' Nucl.\ Phys.\ B {\bf 305}, 219 
(1988).\\

\bibitem{FF}
B. Feigen and E. Frenkel, ``Representations of Affine Kac-Moody Algebras,
Bosonization and Resolutions'', Letters in Mathematical Physics {\bf 19},
307 (1990).\\

\bibitem{FSS}
J.~Fuchs, A.~N.~Schellekens and C.~Schweigert, 
``A Matrix S for all Simple Current Extensions,'' Nucl.\ Phys.\ B {\bf 473}, 
323 (1996) [arXiv:hep-th/9601078].\\

\bibitem{GG}
M.~R.~Gaberdiel and P.~Goddard, ``Axiomatic Conformal Field Theory,'' 
Commun.\ Math.\ Phys.\ {\bf 209}, 549 (2000) [arXiv:hep-th/9810019].\\

\bibitem{TG}
T.~Gannon, ``Boundary Conformal Field Theory and Fusion Ring Representations,'' Nucl.\ Phys.\ B {\bf 627}, 506 (2002) [arXiv:hep-th/0106105].\\

\bibitem{Berk} N. Berkovits, ``An Alternative String Theory in
Twistor Space for N = 4 Super-Yang-Mills'',
Phys.\ Rev.\ Lett.\  {\bf 93}  (2004) 011601[arXiv:hep-th/0402045].\\

\bibitem{BW} E. Witten and N. Berkovits, ``Conformal Supergravity in
Twistor-String Theory'',
JHEP {\bf 0408}, 009 (2004) [arXiv:hep-th/0406051].\\

\bibitem{Wit} E. Witten, ``Perturbative Gauge Theory A String Theory
In Twistor Space'', Commun.\ Math.\ Phys.\  {\bf 252} (2004) 189;
[arXiv:hep-th/0312171].\\


\bibitem{MS} L. Mason and D. Skinner, ``Heterotic Twistor String Theory'',
[arXiv:0708.2276].\\ 

\bibitem{GO} P. Goddard and D. Olive, ``Kac-Moody and Virasoro Algebras in Relation to Quantum Physics'', Int.\ J.\ Mod.\ Phys.\  {\bf A1} (1986).\\

 

\bibitem{CM}H.S.M. Coxeter and W.O.J. Moser, {\sl Generators and Relations for Discrete Groups}, 4th edition, Springer-Verlag, New York, 1980.\\


\bibitem{HTF}A. Erdelyi, ed., {\sl Bateman Manuscript Project: Higher Trancendental Functions},vol. 2,  McGraw-Hill, New York, 1953.\\

\bibitem{RM}R. Miranda, {\sl Algebraic Curves and Riemann Surfaces}, American Mathematical Society, Providence, Rhode Island,  1995.\\

\bibitem{WW}
E.T. Whittaker and G.N. Watson, {\sl A Course of Modern Analysis}, 4th edition, Cambridge University Press, Cambridge, 1927.\\

\bibitem{VSV}
V.S. Varadarajan, {\sl Lie Groups, Lie Algebras and their Representations}, Springer-Verlag, New York,  1984.\\

\bibitem{AWK}A.W. Knapp, {\sl Lie Groups Beyond an Introduction}, Birkh\"auser, Boston, 1996.\\

\bibitem{JEH}
J.E. Humphreys, {\sl Introduction to Lie Algebras and Representation Theory}, Springer-Verlag, New York,  1972.\\

\bibitem{FH}
W. Fulton and J. Harris, {\sl Representation Theory}, Springer-Verlag, New York,  1991.\\

\bibitem{GW}R. Goodman and N.R. Wallach, {\sl Representations and Invariants of the Classical Groups}, Cambridge University Press, Cambridge, 1998.\\

\bibitem{DEL}D.E. Littlewood, {\sl The Theory of Group Characters and Matrix Representations of Groups}, 2nd edition, Oxford University Press, Oxford, 1950.\\





\end{thebibliography}
\end{document}